\documentclass[lettersize,journal]{IEEEtran}
\usepackage{amsmath,amsfonts}
\usepackage{algorithmic}
\usepackage{algorithm}
\usepackage{array}
\usepackage[caption=false,font=normalsize,labelfont=sf,textfont=sf]{subfig}
\usepackage{textcomp}
\usepackage{stfloats}
\usepackage{booktabs}
\usepackage{multirow, threeparttable}
\usepackage{url}
\usepackage{verbatim}
\usepackage{graphicx}
\usepackage{cite}
\usepackage{hyperref}
\usepackage{footnote}
\usepackage{xspace}
\usepackage[table]{xcolor}

\begin{document}

\title{QiMeng: Fully Automated Hardware and Software Design for Processor Chip}

\author{
  Rui Zhang\textsuperscript{1},
  Yuanbo Wen\textsuperscript{1},
  Shuyao Cheng\textsuperscript{1},
  Di Huang\textsuperscript{1},
  Shaohui Peng\textsuperscript{2},
  Jiaming Guo\textsuperscript{1},\\
  Pengwei Jin\textsuperscript{1},
  Jiacheng Zhao\textsuperscript{1},
  Tianrui Ma\textsuperscript{1},
  Yaoyu Zhu\textsuperscript{1},
  Yifan Hao\textsuperscript{1},
  Yongwei Zhao\textsuperscript{1},
  Shengwen Liang\textsuperscript{1},\\
  Ying Wang\textsuperscript{1},
  Xing Hu\textsuperscript{1},
  Zidong Du\textsuperscript{1},
  Huimin Cui\textsuperscript{1},
  Ling Li\textsuperscript{2,3},
  Qi Guo\textsuperscript{1},
  Yunji Chen\textsuperscript{1,3,*}\thanks{* Corresponding Author: cyj@ict.ac.cn}\\
  ~\\
  \textsuperscript{1}State Key Lab of Processors, Institute of Computing Technology, CAS\\
  \textsuperscript{2}Intelligent Software Research Center, Institute of Software, CAS\\
  \textsuperscript{3}University of Chinese Academy of Sciences\\  
  ~\\
  {\url{https://qimeng-ict.github.io/}}
}

\maketitle

\begin{abstract}
Processor chip design technology serves as a key frontier driving breakthroughs in computer science and related fields. With the rapid advancement of information technology, conventional design paradigms face three major challenges: the physical constraints of fabrication technologies, the escalating demands for design resources, and the increasing diversity of ecosystems. Automated processor chip design has emerged as a transformative solution to address these challenges. While recent breakthroughs in Artificial Intelligence (AI), particularly Large Language Models (LLMs) techniques, have opened new possibilities for fully automated processor chip design, substantial challenges remain in establishing domain-specific LLMs for processor chip design. 

In this paper, we propose QiMeng, a novel system for fully automated hardware and software design of processor chips. QiMeng comprises three hierarchical layers. In the bottom-layer, we construct a domain-specific Large Processor Chip Model (LPCM) that introduces novel designs in architecture, training, and inference, to address key challenges such as knowledge representation gap, data scarcity, correctness assurance, and enormous solution space. In the middle-layer, leveraging the LPCM's knowledge representation and inference capabilities,  we develop the Hardware Design Agent and the Software Design Agent to automate the design of hardware and software for processor chips. Currently, several components of QiMeng have been completed and successfully applied in various top-layer applications, demonstrating significant advantages and providing a feasible solution for efficient, fully automated hardware/software design of processor chips. Future research will focus on integrating all components and performing iterative top-down and bottom-up design processes to establish a comprehensive QiMeng system.

\end{abstract}


\section{Introduction}

As the fundamental hardware platform for computing systems, processors and chips undertake critical functions including instruction execution, data processing, and resource management. These processors and chips power diverse devices ranging from personal computers, servers, smartphones, and Internet of Things (IoT) equipment, forming the technological foundation of modern digital economies. Processor chip design represents both a strategically important industry for national economic development and a cutting-edge research field that drives progress in computer science. As a highly complex and systematic task, processor chip design requires tight hardware-software co-design to achieve functional requirements, along with optimizing performance, power, and area (PPA). These requirements make processor chip design one of the most challenging research topics across both industrial and academic domains.

The evolution of information technology has revealed three fundamental limitations in current processor chip design methodologies: constrained fabrication technological, limited resource, and diverse ecosystem. 
In the fabrication technological aspect, as semiconductor fabrication nears physical limits below 7nm nodes, phenomena such as quantum tunneling and short-channel effects become increasingly problematic, rendering conventional fabrication technology-based performance scaling ineffective, thereby necessitating design methodology innovations. 
From a resource perspective, conventional design flows demand extensive expertise and labor-intensive design-verification iteration to ensure functional correctness while balancing competing design objectives such as PPA. This results in protracted development timelines and substantial costs. 
In the ecosystem aspect, emerging applications in Artificial Intelligence (AI), cloud, and edge computing require specialized architectures with customized foundational software support. Thus, conventional chip design approaches cannot meet the ecosystem challenge efficiently due to their inherent lengthy time and substantial cost requirements. To sum up, these challenges underscore the urgent need for novel design paradigms that can deliver enhanced performance, improved efficiency, and reduced costs while meeting diverse application requirements.

Automated processor chip design, which aims to automate the entire design and verification pipeline of processor chips, presents a promising solution to overcome the above-mentioned limitations. By leveraging AI methodologies, automated processor chip design exhibits the potential to surpass manual design and achieve better performance under identical fabrication technology. Additionally, the automated processor design approach is capable of dramatically reducing manual intervention, significantly improving design efficiency while shortening development cycles and lowering costs. Furthermore, it enables rapid customization of chip architectures and software stacks tailored to specific application domains, addressing the growing demand for specialized computing solutions.

Recent breakthroughs in Large Language Models (LLMs) and Multi-Agent systems have created new opportunities for automated processor chip design. State-of-the-art LLMs such as DeepSeek-V3~\cite{dsv3}, DeepSeek-R1~\cite{guo2025deepseek}, Qwen3 \cite{qwen-3}, GPT-4o \cite{gpt4o}, and Gemini 2.5 Pro~\cite{gemini2.5-pro} have demonstrated remarkable capabilities in question answering, planning, and reasoning, exhibiting the potential of artificial general intelligence (AGI). After post-training on domain-specific data, domain-specialized LLMs can be obtained and have shown impressive results across scientific disciplines such as computational biology ~\cite{zhang2025scientific}, materials science, and chemistry~\cite{jablonka202314}. More advanced LLM-based agents integrate cognitive abilities with the tool-use skill of LLMs to autonomously plan and execute complex workflows \cite{wang2024survey}. These developments of LLMs and agents suggest new pathways toward fully automated processor chip design.

Nevertheless, due to the distinctive nature of processor chip design, applying LLMs and agents to automated processor chip design faces four principal challenges: knowledge representation gap, data scarcity, correctness guarantee, and enormous solution space. 
First, the knowledge representation gap: critical processor chip design data employs graph structures, such as abstract syntax trees (ASTs), data flow diagrams (DFGs), and control flow diagrams (CFGs). Graph data exhibits an inherent semantic gap with the sequential text that LLMs typically process, constraining the capacity for domain knowledge representation and limiting the processor chip design capabilities of LLMs. 
Second, the data scarcity: unlike the vast petabyte-scale text corpora available on the Internet for training general-purpose LLMs, processor chip design data are orders of magnitude smaller, with merely terabyte-scale in open-source communities like GitHub, severely constraining the development of domain-specialized LLMs for processor chip design. 
Third, the correctness guarantee: processor design demands rigorous verification standards, which fundamentally conflict with the probabilistic nature of LLMs. For example, Intel's Pentium 4 processor required 99.99999999999\% accuracy in functional verification \cite{li2022competition}.
Finally, the enormous solution space: processor design spans multiple abstraction stages from foundational software to physical layouts, thus, modeling the design space directly at the raw bitstream level suffers from a dimensionality explosion. For example, the solution space for a 32-bit CPU reaches $10^{10^{540}}$. This enormous solution space poses extreme challenges for deriving both functionally-correct and performance-optimized processor designs.

\begin{figure}[t]
    \centering
    \includegraphics[width=1\linewidth]{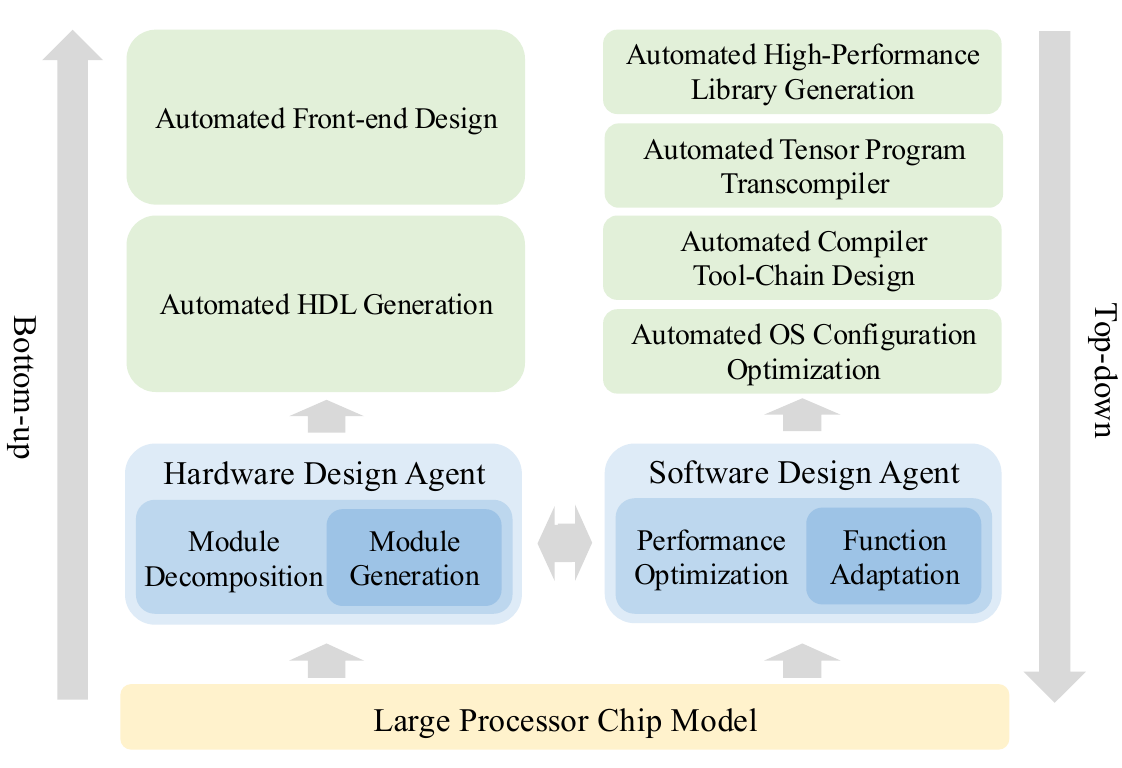}
    \caption{ Overview. QiMeng consists of three layers,  a domain-specialized Large Processor Chip Model (LPCM)  in the bottom-layer, Hardware Design Agent and Software Design Agent enabling automated hardware and software design based on LPCM in the middle-layer, and various processor chip design applications in the top-layer.}
    \label{fig:overview}
\end{figure}

To address the aforementioned challenges and pioneer a transformative paradigm, we propose QiMeng\footnote{QiMeng is a Chinese term that refers to the process of imparting fundamental knowledge and skills to beginners, serving as the cornerstone for intellectual development and skill enhancement. Named by QiMeng, we expect this system can achieve fully automated processor chip design through learning human knowledge and experience, followed by practicing and self-evolving.}, a novel system for fully automated hardware and software design for processor chips. Consisting of three layers, QiMeng constructs a Large Processor Chip Model (LPCM) as a domain-specialized LLM for processor chip design in the bottom-layer and further creates both Hardware Design Agent and Software Design Agent based on LPCM in the middle-layer, enabling automated hardware and software design, respectively.  Finally, the two agents support various processor chip design applications in the top-layer, as shown in Figure \ref{fig:overview}.

In QiMeng, to overcome the above-mentioned four challenges, LPCM is meticulously designed to incorporate domain-specialized knowledge and fundamental competencies of processor chip design. LPCM distinguishes itself from general-purpose LLMs through unique innovations in its architecture, training, and inference. 
Regarding architecture, LPCM employs a multi-modal structure, enabling the comprehension and representation ability of graph data inherent to the processor chip domain, which addresses the critical challenge of the knowledge representation gap. 
For training, it is critical to automatically generate extensive domain-specific data of processor chip design. For each abstraction stage of processor chip design, domain-specific data is systematically collected, and single-stage automated design models are independently trained. These models are subsequently cascaded to autonomously generate extensive cross-stage aligned data for processor chip design. Leveraging this aligned data, LPCM can be trained to learn domain knowledge from the hierarchical design process, effectively mitigating the data scarcity challenge. 
During inference, two feedback-driven mechanisms are implemented. By constructing correctness feedback from automated functional verification, LPCM is able to autonomously repair erroneous results and ensure the validity of generated outputs, addressing the challenge of ensuring correctness in processor design. Concurrently, leveraging performance feedback from automated performance evaluation, LPCM is capable of decomposing the solution space and pruning the low-performance subspaces. Thus, LPCM can effectively reduce the dimensionality of the solution space and enable efficient exploration of high-performance design solutions, overcoming the challenge of the enormous solution space.

Based on LPCM, QiMeng develops two specialized agents, a Hardware Design Agent and a Software Design Agent, dedicated to the automated design of hardware and software for processors and chips.
The Hardware Design Agent adopts a dual-loop mechanism, consisting of an outer module decomposition feedback loop based on performance optimization and an inner module generation feedback loop empowered by automated verification and repair. This dual-loop mechanism facilitates end-to-end automated design from functional specifications to physical layouts, unifying conventional disjointed stages such as logic design, circuit design, and physical design. Thus, Hardware Design Agent enables a fully integrated, cross-stage collaborative design paradigm that is expected to surpass conventional human design, potentially achieving superior performance under identical fabrication technology.
Meanwhile, the Software Design Agent also employs a dual-loop mechanism, consisting of an outer performance optimization feedback loop guided by LLM and an inner function adaptation feedback loop based on automated verification and repair.  Software Design Agent autonomously achieves seamless functional adaptation and performance optimization of foundational software for target processor chips, addressing the dynamic and escalating demands of modern applications.
 
Leveraging the Hardware Design Agent and Software Design Agent, various applications can be developed to address diverse real-world use cases of processor chip design. 
For automated hardware design, significant milestones have been accomplished, including automated front-end design and automated HDL generation. In automated software design, achievements include automated OS configuration optimization, automated compiler tool-chain design, automated tensor program transcompiler, and automated high-performance library generation. These applications have driven the implementation of key components within QiMeng, establishing a solid foundation for its full realization. 
Moving forward, we will construct QiMeng through a three-phase approach, transitioning from top-down to bottom-up, ultimately achieving a self-evolving framework.
Initially, in the top-down phase, the implementation of diverse automated design applications in top-layer will provide two agents in middle-layer with design expertise and generate extensive domain-specific data to enhance the capabilities of the underlying LPCM.
Subsequently, in the bottom-up phase, the improved LPCM, the hardware and software design agents will be applied across a broader spectrum of processor chip design applications in a bottom-up fashion.
Ultimately, in the iteration phase, an iterative cycle integrating top-down and bottom-up approaches will be established to enable the self-evolution of  QiMeng, progressively advancing its fully automated processor chip design capabilities while extending its applicability to support increasingly diverse and complex scenarios. 

Aiming to present a comprehensive framework for fully automated hardware and software design for processor chips, this work introduces QiMeng, along with its roadmap, design methodology, and applications. This paper is structured as follows: Section \ref{sec2} provides the motivation of QiMeng and its roadmap; Section \ref{sec3} elaborates the design of LPCM, encompassing architecture, training and inference; Section \ref{sec4} details the Hardware Design Agent and Software Design Agent; Section \ref{sec5} showcases diverse applications enabled by key components of QiMeng; Section \ref{sec6} surveys related research in automated processor chip design; Section \ref{sec7} concludes with insights into future research trajectories.
\section{Roadmap}\label{sec2}
Automatic processor chip design is one of the central problems in the field of computer science, originating from the Church’s Problem \cite{church1963application}: \textit{How can circuits be automatically designed to satisfy the relationship between given inputs and outputs?}  Proposed in 1957 by Alonzo Church, the founding figure of computer science, this problem has been a major challenge for decades, attracting extensive research from Turing Award winners such as Rabin, Scott, and Pnueli, yet it remains unsolved. 
Early Electronic Design Automation (EDA) tools, which were based on predefined rules and Boolean logic, automated specific design tasks such as logic synthesis, placement, and routing. 
As circuit complexity increased, optimization-based techniques emerged, including High-Level Synthesis (HLS), which automated the translation of high-level descriptions to RTL, and Design Space Exploration (DSE), which optimized design parameters for PPA. 
In recent years, AI technologies have propelled automatic processor chip design into a more intelligent, data-driven phase. Techniques like Random Forests, Reinforcement Learning (RL), and Graph Neural Networks (GNNs) have enabled automatic circuit optimization, placement, and routing, significantly enhancing design efficiency in complex scenarios. However, these approaches mainly apply AI as a tool to refine steps in the conventional EDA process, without fundamentally altering the overall design paradigm.

The current automated design methods have three main limitations. 
First, processor chip design requirements in real-world applications are often expressed in vague, informal natural language, while existing methods can only handle precise, formal inputs, typically in the form of Hardware Description Languages (HDLs). As a result, the transition from informal to formal requires significant work from experts. 
Second, these methods can only automate certain steps of processor chip design, such as logic synthesis, formal verification, automatic placement, and routing. However, critical tasks like logic design, instruction set extensions, software tool-chain adaptation, and optimization still cannot be fully automated. 
Finally, existing methods are typically limited to individual tasks, with a constrained design space and a lack of cross-stage hardware-software co-design, making it difficult to push the boundaries of human-driven design.

The development of LLMs and agents has opened up new possibilities for overcoming three key limitations of the conventional automatic design methodologies. 
First, LLMs can convert informal natural language descriptions into formal programming languages, allowing them to automatically generate correct code for tasks ranging from basic functions to entire programs based on natural language specifications. 
Second, agents built on LLMs can autonomously plan and execute complex tasks and can independently utilize external tools. This capability offers a novel approach for integrating AI techniques with domain-specific tools, which offers new perspectives to achieve fully automated processor chip design. 
Finally, LLMs possess powerful multi-task abilities and demonstrate strong potential in completing complex planning and reasoning tasks, which form the basis for achieving cross-stage collaboration in hardware-software design.

Based on the above analysis, we introduce QiMeng, an innovative paradigm for fully automated hardware and software design for processor chips. 
QiMeng consists of three hierarchical layers, as illustrated in Figure \ref{fig:overview}. The bottom-layer is LPCM, which embeds domain-specialized knowledge in the field of processor chip design. The middle-layer is the Hardware Design Agent and Software Design Agent, which enable the automated design of hardware and software by leveraging the domain knowledge from LPCM. The top-layer focuses on implementing various applications that use the automated design capabilities provided by the Hardware Design Agent and Software Design Agent to address different design requirements for processor chips. 
These three layers work synergistically, forming a complete system for fully automated hardware and software design of processor chips.

However, the realization of QiMeng is not achieved instantaneously. 
Each of the three levels faces its own unique set of challenges, making it difficult to directly establish the complete QiMeng system in a bottom-up way. Specifically, the implementation of LPCM in bottom-layer requires substantial domain-specialized data in hardware/software design of processor chips. However, the domain-specialized data is extremely scarce, preventing the training of LPCM. In the middle-layer, the development of Hardware/Software Design Agent depends on the domain knowledge provided by LPCM, while also needing to integrate specialized tools for verifying the correctness and evaluating performance. At the top-layer, implementation of the various applications relies on both LPCM and the two agents. 
Despite these challenges, the three layers exhibit strong interdependence and can provide mutual enhancement. The various applications at the top-layer can provide valuable domain-specialized data for LPCM, and also facilitate the use of specialized tools for functionality verification and performance assessment towards Hardware/Software Design Agent. Furthermore, achieving constructing a complete interaction process between LPCM and specialized tools, the Hardware/Software Design Agent can offer an automatic data generation mechanism for LPCM. Through the collaborative synergy of the three levels, the challenges each level faces can be effectively resolved.

Although QiMeng is originally designed in a bottom-up manner, it is easier to start with a top-down manner during actual implementation. Driven by the aim of achieving various hardware and software designs, the implementation of applications in top-layer can offer extensive synthetic domain-specialized data for LPCM, and also provide design experience of collaborating with specialized tools for designing Hardware/Software Design Agent.

Based on the above analysis, we propose a three-phase roadmap to implement a complete QiMeng system. 
The first phase is to adopt a top-down construction approach, developing various applications based on the LPCM, which is initialized with a general-purpose LLM.
During the implementation of applications, key components of and functions of the Hardware/Software Design Agent are constructed, which are then combined to establish complete processes of the two agents. At the same time, extensive domain-specialized data of software and hardware design is synthesized for training LPCM, enabling LPCM to acquire domain knowledge superior to general-purpose LLMs. 
The second phase is to adopt a bottom-up construction approach, reconstructing the Hardware/Software Design Agent based on the trained LPCM and re-developing the various applications. Due to LPCM being enhanced with domain knowledge and specialized capabilities, the applications redeveloped in the second phase will achieve better automated design results than those in the first phase. On this basis, higher-quality software and hardware design data can be obtained based on the applications in the second phase. 
The third phase is to form an iterative loop by combining the top-down and bottom-up design processes. Inspired by John von Neumann's ``Theory of Self-Replicating Automata'' \cite{vonNeumann1966}, we hope to achieve the self-evolution of QiMeng through this loop. In the process of evolution, on the one hand, we aim to expand the depth of QiMeng, continuously improving its capabilities for fully automated software and hardware design for processor chips. On the other hand, we aim to expand the breadth of  QiMeng, continuously extending the spectrum of applications and providing intelligent support for a broader range of processor chip design scenarios.

The current work is still in the first phase of the three-phase approach. So far, representative applications including automated front-end design, automated HDL  generation, automated OS configuration optimization, automated compiler tool-chain design, automated tensor program transcompiler, and automated high-performance library generation have been successfully implemented. These applications fulfill some key components of the Hardware/Software Design Agent. In future work, we will complete the first phase of integrating these key components into a complete Hardware/Software Design Agent and automatically generate extensive domain-specialized data to train LPCM. Following this, the second and third stages will be carried out to build a comprehensive QiMeng system.

\section{Large Processor Chip Model}\label{sec3}

\begin{figure*}[t]
    \centering
    \includegraphics[width=\linewidth]{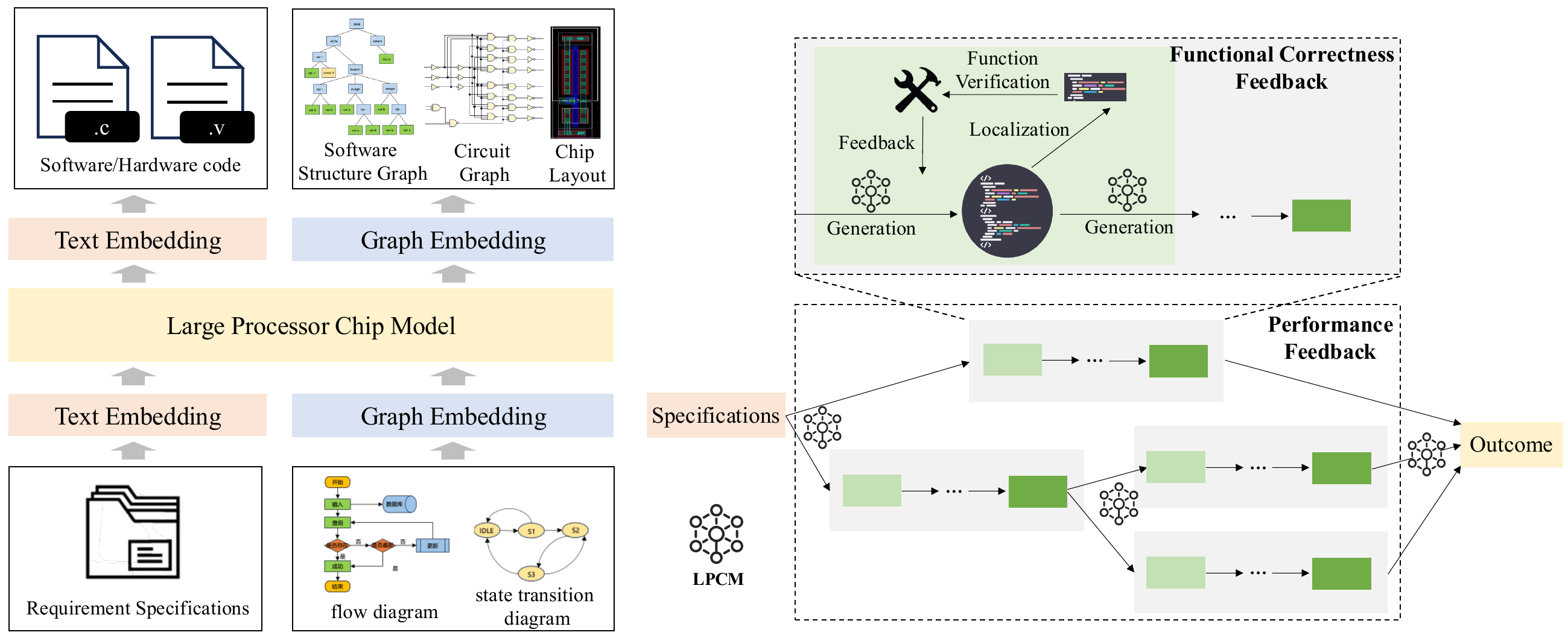}
    \caption{Left: Multimodal architecture of LPCM capable of understanding, representing, and generating both text and graph data. Right: Feedback-driven inference of LPCM with a dual-loop mechanism, consisting of an outer performance feedback loop and an inner functional correctness feedback loop.}
    \label{fig:LPCM}
\end{figure*}

Due to the unique nature of the processor chip design, four key challenges must be addressed: knowledge representation gap, data scarcity, correctness guarantee, and enormous solution space. To tackle these challenges, LPCM developed in QiMeng employs distinctive approaches in architecture, training, and inference, setting it apart from general-purpose LLMs. This section provides a detailed discussion of these innovations.

Notably,  designing a comprehensive LPCM immediately poses significant challenges, as it demands expertise and capabilities of foundational software development to chip design. Our prior work \cite{chang2025largeprocessorchipmodel} investigates the construction of LPCM, categorizing the process into three levels: 1) Human-Centric, which assists and provides suggestions for humans in code generation and parameter tuning; 2) Agent-Orchestrated, independently completing certain subtasks with toolchain integration to facilitate cross-layer optimization; 3) Model-Governed, achieving full automation of the full process of hardware-software co-design, simulation, and iterative refinement. This paper specifically focuses on the design methodology of level 3.

\subsection{Multimodal Structure}
Since text is typically organized as sequential data, most existing LLM architectures are primarily focused on handling sequential information. Even in multimodal LLMs that process images or data of other modalities, multimodal features are treated as a specialized sequence type and are concatenated with text sequences before feeding into the model. 
However, in the processor chip design domain, beyond textual descriptions of functional requirements and formal code representations, much of the critical information and knowledge is represented as graph illustrations. For example, software architecture is often represented as ASTs, chip logic architecture as DFGs, and chip circuit architecture as CFGs. 
These graph data are essential for processor chip design. As a result, LPCM is specifically designed as a multimodal architecture, capable of understanding, representing, and generating graph data, enabling more effective capacities of learning and presenting processor chip domain knowledge, as shown in Figure \ref{fig:LPCM} left.

Specifically, the input to LPCM consists of two modalities: textual descriptions and graphical illustrations of requirement specifications. There are two critical issues in understanding and representing graph data: feature representation and feature alignment. 
A straightforward approach is to represent the graph data in a special textual format and concatenate it with the textual tokens before feeding it into the model. However, this approach serializes the graph’s topological structure, potentially causing nodes that are topologically close in the graph to be positioned far apart in the sequence, thus losing topological information of the graph data. 
To better preserve the graph's topological information, Graph Neural Networks (GNNs) \cite{GNNMScarselli} can be used to encode the graph data and generate its embedding. Contrastive learning can then be applied to align the features of the graph embedding with the corresponding text embedding. Once feature alignment is achieved, the graph embedding is concatenated with the other textual tokens and fed into LPCM.

The output of LPCM also encompasses two modalities: text and graph. The text modality includes both the software and hardware code for processor chip design, while the graph modality includes generated diagrams, such as software architecture diagrams, chip circuit diagrams, and chip layouts. 
The process of generating a graph is closely tied to its representation. If the graph data is directly represented by a special textual format, LPCM can also directly output the graph in this format. However, this representation method may risk losing topological information, which could compromise the accuracy of the generated graphs. 
To better preserve the topological information of the graph, LPCM can first output the graph's embedding, which is then mapped to a graph structure using specialized graph generation models, such as diffusion models like GRAPHARM \cite{kong2023autoregressive} or generative GNNs like GPT-GNN \cite{hu2020gpt}. To more accurately reflect the characteristics of the circuit, the Binary Decision Diagram (BDD), which is commonly used in the context of circuit design, can be employed. Additionally, Binary Speculation Diagram (BSD) \cite{cheng2024automated}, which is an enhanced version of BDD for circuit generation, can also be used to offer better suitability for automated processor chip design.

\subsection{Cross-stage Collaborative Training}
To enable the automated full-process design of processor chips, it is essential to gather cross-stage design data for training LPCM. 
However, the processor chip field faces significant data scarcity. In comparison to the petabyte-scale text corpora available on the Internet, software and hardware code data from open-source communities like GitHub are limited to terabyte-scale. Moreover, this data typically covers only specific stages of the processor chip design process. Thus, cross-stage design data, which includes multiple stages of design abstraction, are seriously scarce. This data scarcity challenge presents a major obstacle to the effective training of LPCM. 
Therefore, it is crucial to develop a cross-stage collaborative design database to provide the necessary foundation for training LPCM.

To build a cross-stage collaborative design database, an automated process for generating cross-stage design data should be established. 
For multiple abstract stages of processor chip design, including high-performance library design, OS kernel design, compiler tool-chain design, logic design, circuit design, and physical design, design data is first collected separately for each stage. These data only need to capture information for individual stages and do not require cross-stage alignment. Higher-level information tends to resemble natural language and contains richer semantic content, while lower-level information is closer to the actual graphical representation of the chip. Therefore, data at each stage must include both textual and graphical modalities. 
These single-stage design data are then used to train models, which generate automated design models for each stage. By cascading these models together, large-scale, cross-stage aligned hardware and software design data for processor chips can be automatically generated to address the challenge of data scarcity.

Once the cross-stage collaborative design database is constructed, we can train LPCM to develop the capability to generate cross-stage collaborative design reasoning. This can be achieved by applying Chain-of-Thought (CoT) imitation learning based on the database. In this process, design data from multiple stages of the processor chip design in the cross-stage collaborative design dataset are treated as reasoning sequences, generating numerous (input, CoT, output) triplets, which are used to train the CoT reasoning of LPCM. To enhance the LPCM's understanding of intermediate stage design details, a distribution alignment loss is introduced into the training objective. 
To ensure stable training, a curriculum learning strategy can be employed, starting with samples that feature shorter CoT and simpler design complexities, progressively increasing the complexity of samples. Training on this comprehensive cross-stage collaborative design database will enable the processor model to acquire the capabilities of generating a collaborative design process and the final design outcome. 
Furthermore, an automated unit testing framework will be employed to create a reward function, and RL methods will be applied to further refine the LPCM's CoT generation and improve its processor chip design capabilities.

\subsection{Feedback-Driven Inference}
Although LPCM is equipped with domain-specialized knowledge to handle graph information and capabilities of cross-stage collaborative design, challenges such as context length limitations and hallucinations hinder the ability of LPCM to achieve seamless end-to-end processor chip design and foundational software adaptation and optimization. 
To ensure both the correctness and high-performance of the generated hardware/software of processor chips, it is essential to develop a feedback-driven inference mechanism for LPCM to facilitate effective design planning, leverage external functional verification tools, and optimize performance through feedback. 
Specifically, feedback-driven inference can be divided into two categories: functional correctness feedback and performance feedback, as shown in Figure \ref{fig:LPCM} right.
LPCM perform these two feedback with a dual-loop mechanism, consisting of an outer  performance feedback loop and an inner functional correctness feedback loop.

\subsubsection{Functional Correctness Feedback}
Functional verification is a critical step in ensuring the correctness of manually designed processor chips. By simulating various use cases and corner cases, this process verifies the functionality of key components such as the processor's instruction set, data paths, control logic, and multi-core coordination. Functional verification ensures that the processor chip meets specifications before tape-out, preventing costly re-manufacturing due to logical errors or functional defects. 
Functional verification is essential across all stages of processor chip design, including logic design, circuit design, and physical design. During verification, techniques such as formal verification, dynamic simulation, and hardware simulation are employed. If functional errors are identified, experts manually refine the design and perform iterative verification until the design achieves full functional correctness. 
Inspired by this, to advance functional correctness during the automated processor chip design process, a feedback mechanism oriented to functional Correctness must be integrated into the inference of LPCM. This functional correctness feedback mechanism uses verification feedback to automatically verify the design and further repair the design errors, ensuring the correctness of the design outcomes and addressing the challenge of correctness guarantee.

To implement functional correctness feedback in inference, additional automated verification and repair loops are integrated into the intermediate steps of reasoning. 
Specifically, LPCM actively assesses whether automated verification is necessary for the current reasoning step during inference. When automated verification is required, LPCM utilizes appropriate specialized tools or models to validate the functionality of the intermediate design. 
If a functional error is detected, automated repair is triggered, which involves reverting to the last verified functional step in the reasoning chain, incorporating error feedback from the current validation, and regenerating the design of the current step. This process of verification and repair is repeated iteratively until a functionally correct design is achieved. 
Through this iterative functional correctness feedback loop, the design's correctness is continuously refined, ultimately approaching 100\% functional correctness of the automatically designed processor chips.

\subsubsection{Performance Feedback}
Performance optimization aims to improve the PPA of designed process chips, playing a vital role in designing processor chips and adapting foundational software. 
Existing automated optimization tools, such as deep learning compilers and DSE methods, typically rely on expert-designed rule-based optimization methodologies or machine learning techniques. While these tools significantly reduce the human labor compared to manual tuning, they still encounter issues such as narrow coverage, suboptimal efficiency, and poor cross-domain transferability.
To strengthen the capacity of automated performance optimization, cross-stage collaborative design is necessary. LPCM should be capable of directly generating chip layouts from functional specifications. Nevertheless, formulating the design issue directly with the raw bitstream input of processor chips leads to the curse of dimensionality. For instance, the solution space for a 32-bit CPU could grow to $10^{{10}^{540}}$.

To address the challenges of the enormous solution space, processor chip design has been divided into multiple abstract stages. During logic design, functional specifications are translated into high-level HDL. In circuit design, these high-level HDLs are converted into gate-level netlists. While in physical design, gate-level netlists are turned into chip layouts. 
This workflow follows a progressive coarse-to-fine design process. Each stage incorporates additional implementation details, progressively introducing more constraints, gradually pruning the solution subspaces with lower probability of containing optimal solutions, thereby reducing the overall solution space size.
Inspired by this, to enable automated performance optimization, LPCM needs to adopt a hierarchical search-based inference mechanism guided by performance feedback. This involves building hierarchical decompositions, where the solution space is pruned based on domain knowledge and performance feedback, effectively reducing its dimensionality. 
Simultaneously, by leveraging the iterative reasoning and Test-Time Scaling (TTS) benefits demonstrated by LLMs during inference, the solution space can be efficiently explored, addressing the enormous solution space challenge and enhancing the performance of automated design.

To implement performance feedback in inference, efficient search techniques must be integrated into the inference process to obtain high-performance outcomes from the vast solution space. 
Specifically, LPCM can generate different optimization strategies depending on the target hardware architecture and software characteristics, thus building a search tree in which the initial result is the root node, while intermediate nodes and leaf nodes represent the current optimized outcomes. 
Additionally, by predicting the performance of intermediate nodes and utilizing performance feedback from real-world deployment tests, LPCM can prune suboptimal search branches and generate further optimization strategies based on the current optimal search branch until either reaching a fixed optimization budget or encountering performance improvement saturation.
Through tree search-based inference guided by performance feedback, the performance of hardware and software can be progressively improved, tailored to specific scenarios.
\section{Processor Chip Design Agents}
\label{sec4}

Building upon LPCM, QiMeng develops two specialized agents: Hardware Design Agent and Software Design Agent, to enable fully automated hardware/software design for processor chips.

\subsection{Hardware Design Agent}

With fabrication technologies approaching physical limits, the conventional approach of enhancing chip performance through process scaling below 7nm nodes has encountered fundamental limitations. Therefore, the focus needs to pivot from fabrication-centric advancements to design methodologies innovations. 
To overcome the fabrication technological constraint in processor chip design, it is necessary to explore fully automated hardware design. This involves establishing a seamless design framework from functional specifications to physical layout, bypassing the conventional multi-stage design hierarchy, including logic design, circuit design, and physical design, to explore a broader cross-stage collaborative design space for superior solutions.   
The automated design framework must fulfill two critical objectives: 1) Functional correctness, guaranteeing that the automatically designed processor chip delivers accurate computational outcomes; and 2) High performance, optimizing performance such as computational throughput, power efficiency, and area utilization.

\begin{figure*}[t]
    \centering
    \includegraphics[width=1\linewidth]{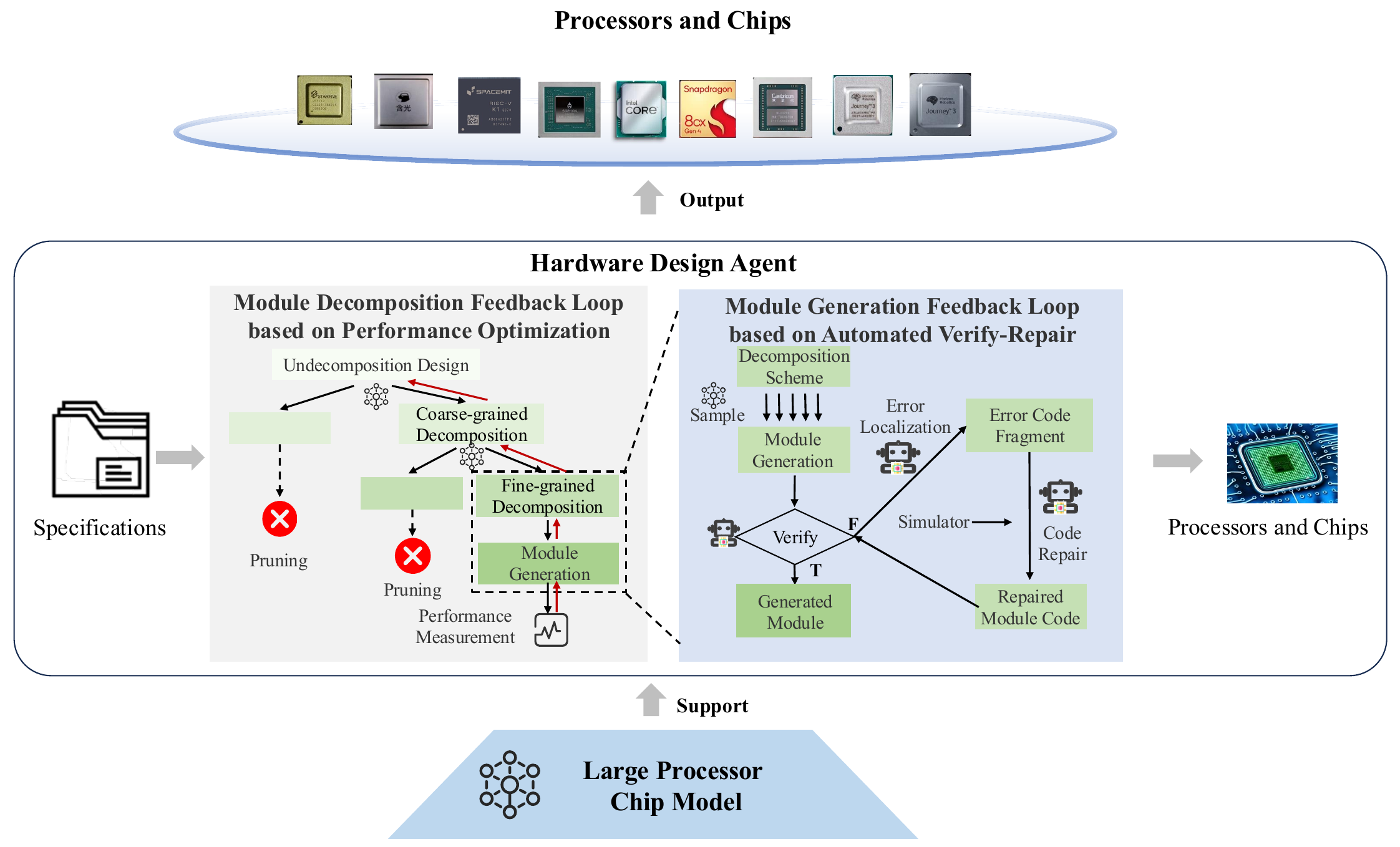}
    \caption{Structure of Hardware Design Agent which features a dual-loop feedback mechanism: an outer module decomposition feedback loop based on performance optimization and an inner module generation feedback loop empowered by automated verification and repair.}
    \label{fig:hardware}
\end{figure*}

To accomplish the aforementioned two critical objectives, the Hardware Design Agent needs to implement automated module decomposition and module generation. This enables obtaining a performance-optimized fine-grained design scheme through module decomposition, followed by generating functionally correct modules, ultimately achieving fully automated hardware design results. 
Automatic module decomposition addresses the enormous solution space challenge in hardware design by dividing the processor architecture into functionally independent and verifiable modules, aiming at solution space reduction and global performance optimization. Notably, there exist multiple valid decomposition schemes that satisfy functional specifications but exhibit substantial performance variations. Therefore, it is essential to establish a performance-driven decomposition mechanism to enhance the PPA of processor chips.
Following decomposition, module generation proceeds according to both functional specifications and the selected decomposition scheme. The subsequent integration of these modules yields the final hardware design outcomes. During module generation, functional correctness must be guaranteed. To this end, Hardware Design Agent synergistically combines LPCM with symbolic methods, establishing an automated verification and repair mechanism to ensure the functional correctness of generated modules.
Specifically, LPCM enables the transformation from informal, natural language specifications to formal HDL, accommodating diverse application requirements. After that, using symbolic representations based on BSD \cite{cheng2024automated} to automatically verify and repair the generated module, ensuring functional correctness.

Based on LPCM, we develop a Hardware Design Agent to achieve automated processor chip design from a high-level specification. Hardware Design Agent features a dual-loop feedback mechanism: an outer module decomposition feedback loop based on performance optimization and an inner module generation feedback loop empowered by automated verification and repair, as shown in Figure \ref{fig:hardware}. The outer loop addresses the enormous solution space challenge through module decomposition, while the inner loop addresses the correctness guarantee challenge through automated verification and repair.
In implementation, the outer loop is initiated with the undecomposed design as the root node of the search tree. In each iteration, LPCM proposes and identifies promising finer-grained decomposition candidates that bring potential performance gains as child nodes, based on functional specifications, the current decomposition state, and accumulated domain knowledge. Namely, among the available decomposition schemes, nodes demonstrating suboptimal performance are discarded, which simultaneously reduces the solution space dimensionality and lowers computational complexity. This iterative process constructs a module decomposition search tree where leaf nodes correspond to complete decomposition schemes. The final performance evaluation of each leaf node integrates the outer module decomposition scheme with the inner verified modules, enabling backtracking-based optimization and searching of the module decomposition schemes. The evaluation results of the current module decomposition scheme are added to the domain knowledge base to support subsequent module decompositions.
Simultaneously, in the inner loop, LPCM extracts functional specifications corresponding to the target module based on the module decomposition scheme and then generates the corresponding HDL. However, the initially generated HDL may contain functional inaccuracies. To resolve it, it is necessary to verify and repair the generated modules based on the capacity of functional correctness feedback in the inference of LPCM.
Specifically, the HDL of target module is first transformed into a BSD representation, along with a subset of input-output pairs sampled from the truth table for simulation-based validation. When discrepancies arise, the erroneous BSD nodes undergo Shannon expansion, facilitating automated error correction. Each repair cycle monotonically increases the BSD's functional accuracy. Through repeated verify-repair iterations, the BSD asymptotically approaches 100\% correctness, ultimately producing a validated module design.

\subsection{Software Design Agent}

\begin{figure*}[t]
    \centering
    \includegraphics[width=1\linewidth]{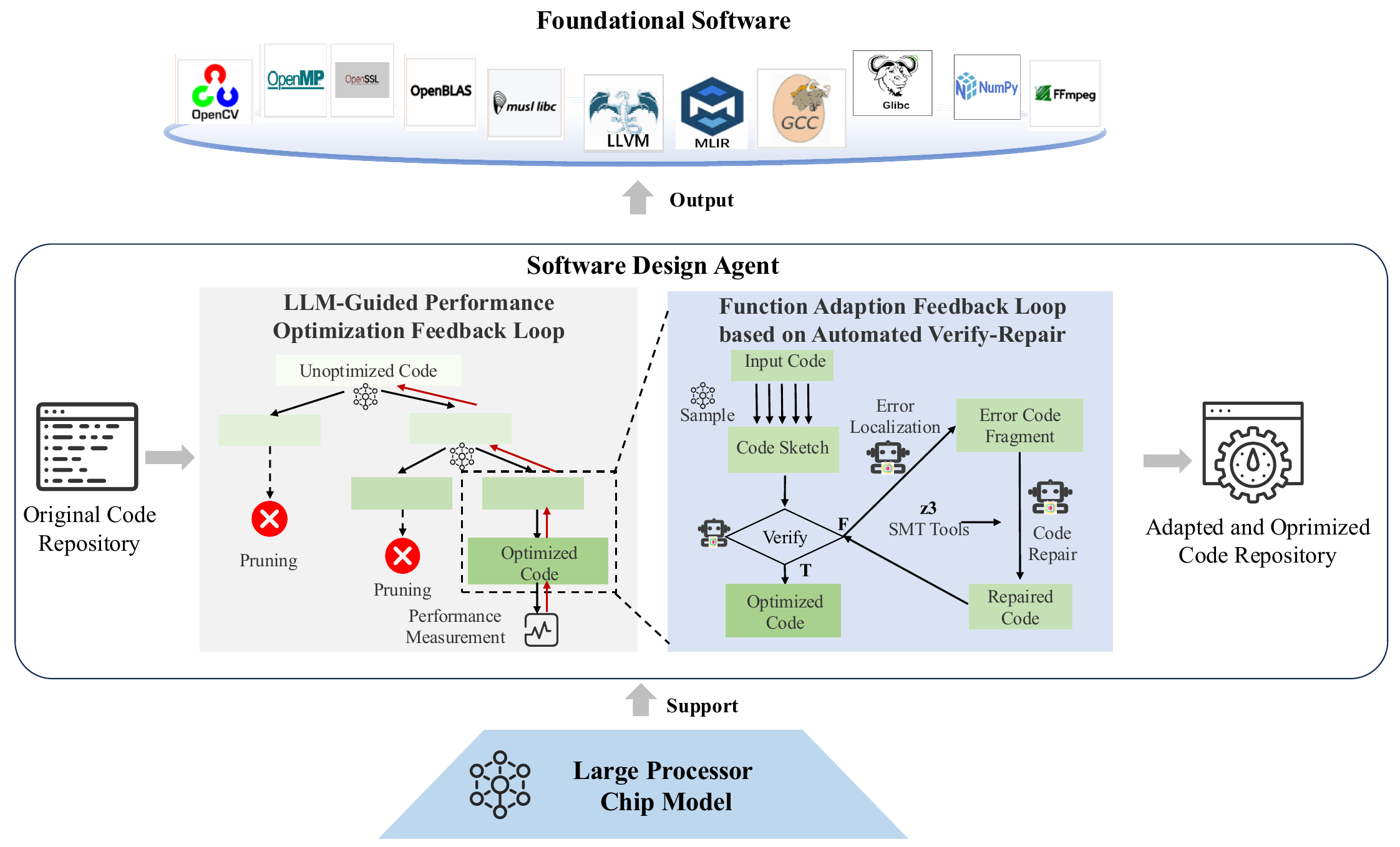}
    \caption{Structure of Software Design Agent which uses a dual-loop feedback mechanism: the outer loop focusing on performance optimization feedback through LLMs-guided search and the inner loop managing function adaptation feedback through automatic verification and repair.}
    \label{fig:software}
\end{figure*}

Foundational software plays a pivotal role in establishing comprehensive technology ecosystems for processor chips, serving as the decisive factor for their successful commercialization and widespread adoption. Nevertheless, adapting and optimizing such software presents formidable obstacles, particularly given the current landscape of fragmented instruction set architectures and diverse software ecosystems. The RISC-V ISA exemplifies this challenge, while its open and modular design offers unprecedented flexibility, it simultaneously introduces complexity orders of magnitude greater than x86/ARM architectures. With nearly 100 optional instruction extensions, including Vector Extension, Matrix Extension, and Cryptographic Extension, the combinations grow exponentially, while each variant demands meticulous compatibility verification across the entire software stack. For example, the openEuler OS \cite{zhou2022openeuler} comprises over 10,000 repositories containing 4 million files, all requiring exhaustive validation for different RISC-V instruction combinations. This combinatorial explosion renders conventional manual development approaches impractical, crystallizing two fundamental challenges: 1) achieving comprehensive functional adaptation to ensure software stability across diverse instruction sets, and 2) conducting deep performance optimization to fully exploit hardware capabilities.

To address these challenges for developing the foundational software ecosystem on specific processors, it is essential to implement AI-driven automated methods for function adaptation and performance optimization of foundational software. The function adaptation problem can be abstracted as a ``program generation'' task, where an agent translates the source code/platform into a target code/platform. Typical applications include automated compiler tool-chain design and automated tensor program transcompilers~\cite{han2022cox,johnson2022martini,papakonstantinou2013efficient,bahdanau2014neural,roziere2020unsupervised,li2023starcoder}. The key to this approach is synergistically combines neural and symbolic methods: LLMs handle high-level program skeleton generation through meta-prompts for flexibility, while SMT-based program synthesis ensures correctness by rectifying low-level implementation errors. 
The performance optimization problem for foundational software can be abstracted as a ``search'' problem where the goal is to efficiently explore the enormous space of optimization primitives and parameter combinations to find the optimal configuration. 
Typical applications include the automated generation of high-performance operator libraries~\cite{zheng2020ansor,bi2023heron} and OS configuration optimization~\cite{chen2024autoos}.
LLMs can efficiently guide this search by leveraging their in-context learning capabilities through carefully designed meta-prompts encoding hardware characteristics and optimization primitives, then effectively pruning the search space to discover optimal implementations tailored to specific hardware-software combinations.

Following this approach, we developed the Software Design Agent based on LPCM, as shown in Figure~\ref{fig:software}. 
The Software Design Agent uses a dual-loop feedback mechanism, with the outer loop focusing on performance optimization feedback through LLMs-guided search and the inner loop managing function adaptation feedback through automatic verification and repair.
This process can finally transform the original code repository into one that is both adapted and optimized to enhance the foundational software ecosystem. 
Note that the outer loop addresses enormous solution space challenges through hierarchical decomposition and optimization feedback, while the inner loop tackles correctness guarantee challenges through automated verification and repair. 
Specifically, in the outer performance optimization feedback loop, the original code is used as the starting point for a Monte Carlo Tree Search~\cite{browne2012survey}. Then, the domain expert knowledge from the LLMs helps evaluate the search tree, prune inefficient branches, and select branches with potential performance gains. The tree is finally refined based on performance measurements, forming an iterative ``observe-prune-optimize-evaluate'' loop until the desired optimization results are achieved. 
In the inner function adaptation feedback loop, we utilize TTS of LLMs for program sampling to generate diverse program sketches, followed by unit testing and execution trace analysis on high-quality sketches to identify minimal erroneous fragments. These fragments are then repaired using solver-based program synthesis such as Z3~\cite{de2008z3}, regarding the original implementation, iterating the ``generate-verify-repair'' loop until functional equivalence is achieved.

\section{Applications}
\label{sec5}

\begin{table}[t]
\centering
\caption{Results of automated front-end design by QiMeng-CPU series compared with existing methods.}
\begin{tabular}{cccc}
     \toprule
                Target Circuit    &  Methods    &Scale   &Performance    \\
     \midrule
 Adder~\cite{roy2021prefixrl} &RL & 118 & NA  \\ 
 Circuit Modules~\cite{chen2020circuit} &DT & 186 & NA   \\ 
 Circuit Modules~\cite{rai2021logic} &EL & $\sim$ 2500 & NA  \\ 
 8-bit CPU~\cite{wu2024chateda} &LLM & 999 & NA  \\ 
 \midrule
QiMeng-CPU-v1  & \multirow{2}{*}{BSD} & \multirow{2}{*}{$\sim$ 4 Million} & \multirow{2}{*}{$1.62\times 10^4$ } \\ 
(RISC V-32 CPU)~\cite{cheng2024automated} & & & \\ 
 QiMeng-CPU-v2  &\multirow{2}{*}{S-BSD}  &\multirow{2}{*}{ \textbf{$\sim $17} \textbf{Million}} & \multirow{2}{*}{$\mathbf{6.29\times10^6}$} \\
(Superscalar CPU)~\cite{cheng2025qimengcpuv2automatedsuperscalarprocessor} & & & \\

     \bottomrule
\end{tabular}
\label{tab:qimeng-cpu}
\end{table}

Leveraging LPCM and Hardware/Software Design Agents,  QiMeng has developed a series of innovative automated design applications to address various hardware/software design requirements. These implementations effectively solve practical needs by strategically applying specific components of Hardware/Software Design Agents and achieving hardware/software automated design for processor chips.

\begin{figure}[t]
    \centering
    \includegraphics[width=1\linewidth]{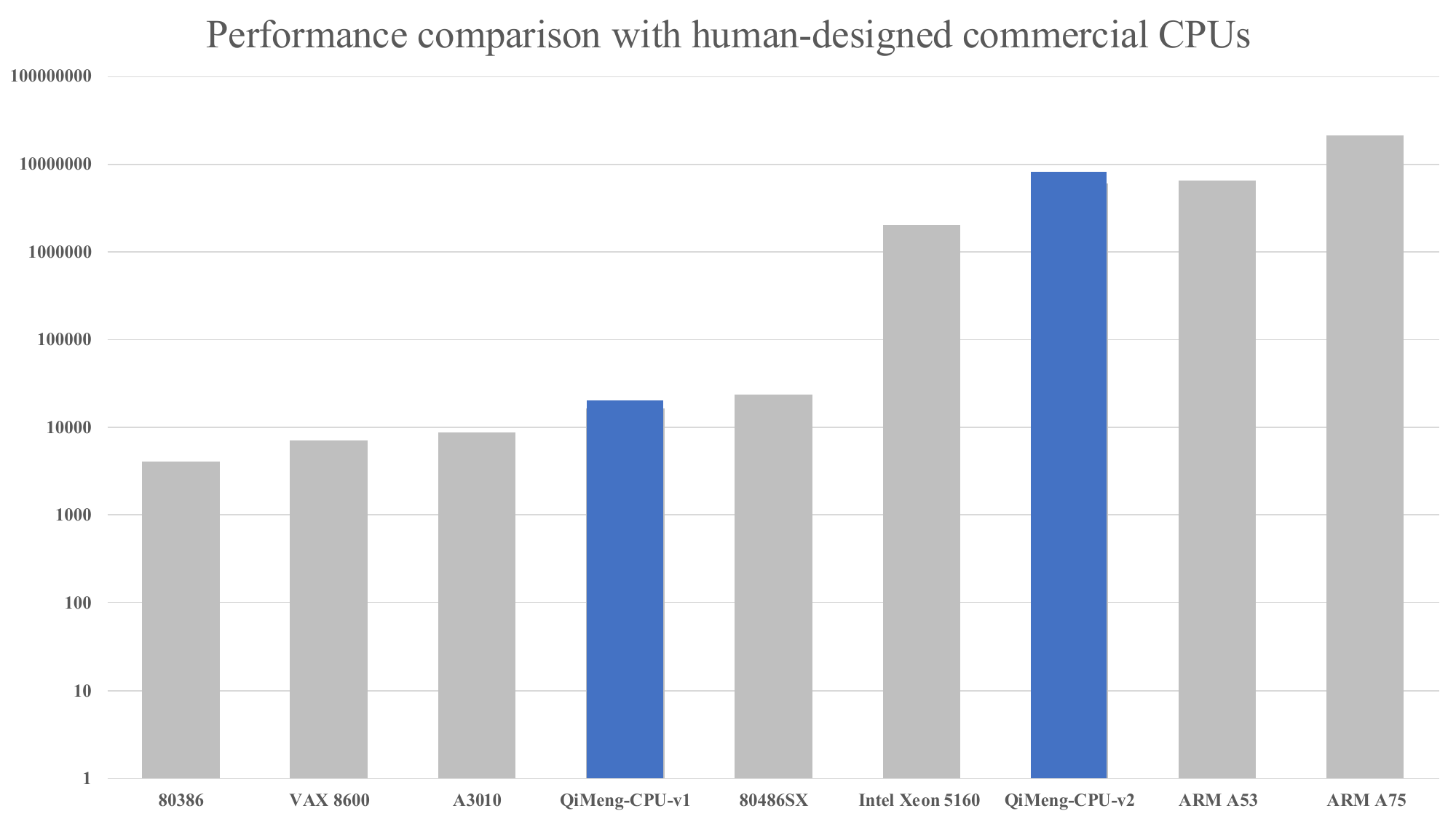}
    \caption{ Performance of QiMeng-CPU series (blue) compared with human-designed commercial CPUs (grey) on the official ARM CPU benchmark Dhrystone. The results show that QiMeng-CPU-v1 is comparable to Intel 486 (1990s CPU), while QiMeng-CPU-v2 is comparable to Arm Cortex A53 (2010s CPU).}
    \label{fig:qimeng-cpu}
\end{figure}

\subsection{Automated Front-end Design}
Automating the design of general-purpose computer CPUs has been a pivotal research challenge since the 1950s, drawing the attention of AI pioneers like Turing and Church \cite{church1963application}. With the advancement of AI technologies, various methods such as decision trees \cite{chen2020circuit}, LLMs \cite{wu2024chateda},  and RL \cite{roy2021prefixrl} have been attempted for automated circuit design. Nevertheless, the absence of well-defined formal representations of circuits limits the precision of existing methods, restricting current capabilities to circuits of roughly thousands of gates without guaranteeing accuracy at larger-scale circuits.

To achieve automated design for large-scale processors and chips, we employ the module generation feedback loop based on automated verification and repair within the Hardware Design Agent to ensure functional correctness, while adopting Binary Speculation Diagrams (BSD) as the circuit's graph-based representation \cite{cheng2024automated}. BSD exhibits two key characteristics for combinational logic circuits: 1) design accuracy improves monotonically with the number of design nodes, and 2) accuracy asymptotically converges to 100\% as the number of data sampling increases. 
The implementation initializes with a randomly generated BSD, based on the automated verification and repair feedback, and iteratively verifies the current BSD in a simulator. When errors are detected, the corresponding BSD nodes are repaired by Shannon expansion, thereby monotonically increasing the functional accuracy of the BSD. By iteratively cycling the automated verification and repair steps, the functional accuracy progressively converges to 100\%. Applying this methodology, the entire front-end design of a 32-bit RISC-V CPU was automatically completed within 5 hours, producing QiMeng-CPU-v1 (also named Enlightenment-1), which is the world's first processor core designed fully automatically \cite{cheng2024automated}. As shown in Table \ref{tab:qimeng-cpu}, QiMeng-CPU-V1 has about 4 million gates, more than 1700$\times$  larger than existing work and achieving a industrial-scale.
Taped out in 2021, QiMeng-CPU-v1 achieves computational performance comparable to Intel's 1990s-era 486 processors, as shown in Figure \ref{fig:qimeng-cpu}.

In addition, we leverage the module decomposition feedback loop based on performance optimization within the Hardware Design Agent to enhance the performance of automated front-end design. For automated pipeline design, gate-level dependency analysis is employed to explore decomposition strategies for pipeline modules, enabling the identification of more efficient fine-grained gate-level pipeline partitioning solutions. Subsequently, a gate-level pipeline controller is implemented, facilitating gate-level short-path forwarding. Finally, the decomposed pipeline modules are synthesized into circuits using BSD. The resulting gate-level pipelines yield an average performance gain of 1.57$\times$, outperforming manually designed counterparts by 37\% in throughput \cite{cheng2024revisiting}. Meanwhile, in the automated design of superscalar processors, a simulated annealing algorithm is applied to search for predictable processor states, enabling instruction-level module decomposition. A Stateful Binary Speculation Diagram (S-BSD) architecture is then devised to generate instruction modules by predicting inter-instruction data dependencies, thereby achieving instruction-level parallelism and enhancing processor performance. Leveraging this methodology, we develop the world’s first automated designed superscalar CPU, QiMeng-CPU-v2 \cite{cheng2025qimengcpuv2automatedsuperscalarprocessor}, which delivers about 380$\times$ speedup over single-cycle predecessor QiMeng-CPU-v1 (Enlightenment-1), and matches the performance of the ARM Cortex A53, as shown in Table \ref{tab:qimeng-cpu} and Figure \ref{fig:qimeng-cpu}.

Notably, the design of QiMeng-CPU-v1 initializes with random circuits and leverages the module generation feedback loop based on automated verification and repair within the Hardware Design Agent, whereas QiMeng-CPU-v2 extends by further integrating the module decomposition feedback loop based on performance optimization within the Hardware Design Agent. However, both of them currently operate without utilizing LPCM. In subsequent research, we intend to integrate the automated HDL generation methods (introduced in Section \ref{sec-HDL}) based on LPCM with the verification-repair-guided module generation loop from QiMeng-CPU-v1 and performance-guided module decomposition loops from QiMeng-CPU-v2, ultimately constructing the complete Hardware Design Agent.

\subsection{Automated HDL Generation}\label{sec-HDL}
\begin{table*}[t]
\caption{Comparison of our CodeV series against various baseline models. Results are cited from the original paper.}
\label{tab:ve1&rtllm1}
\centering
\footnotesize
{%
\begin{tabular}{cccccccccccc}
\toprule
\multirow{2}{*}{Type} & \multirow{2}{*}{Model} & \multirow{2}{*}{\begin{tabular}[c]{@{}c@{}}Model\\ size\end{tabular}} & \multirow{2}{*}{\begin{tabular}[c]{@{}c@{}}Open\\ source\end{tabular}} & \multicolumn{3}{c}{VerilogEval-Machine (\%)} & \multicolumn{3}{c}{VerilogEval-Human (\%)} & \multicolumn{2}{c}{RTLLM v1.1 (\%)} \\ 
 &  & & &   pass@1 & pass@5 & pass@10 & pass@1 & pass@5 & pass@10 &  Syntax & Func. \\
 \midrule
\multirow{7}{*}{Base LLMs} & GPT-3.5 & - & \text{\texttimes} & 60.9 & 75.0 & \multicolumn{1}{c}{79.9} & 33.5 & 45.9 & 50.0 & \multicolumn{1}{c}{79.3} & 51.7 \\
 & GPT-4 & - &\text{\texttimes} & 60.0 & 70.6 & \multicolumn{1}{c}{73.5} & 43.5 & 55.8 & 58.9 & \multicolumn{1}{c}{\textbf{100.0}} & 65.5 \\
 & StarCoder~\cite{starcoder} & 15B & \checkmark & 46.8 & 54.5 & \multicolumn{1}{c}{59.6} & 18.1 & 26.1 & 30.4 & \multicolumn{1}{c}{93.1} & 27.6 \\
 & CodeLlama~\cite{codellama} & 7B & \checkmark & 43.1 & 47.1 & {47.7} & 18.2 & 22.7 & 24.3 & {86.2} & 31.0 \\
 & DeepSeek-Coder~\cite{deepseekcoder} & 6.7B & \checkmark & 52.2 & 55.4 & {56.8} & 30.2 & 33.9 & 34.9 & \multicolumn{1}{c}{93.1} & 44.8 \\
 & CodeQwen~\cite{qwen} & 7B & \checkmark & 46.5 & 54.9 & \multicolumn{1}{c}{56.4} & 22.5 & 26.1 & 28.0 & \multicolumn{1}{c}{86.2} & 41.4 \\
 & Qwen2.5-Coder~\cite{qwen25coder} & 7B & \checkmark & 66.2 & 79.2  & 83.9  & 34.6  & 45.6  & 51.0  & 89.6 & 41.4 \\
 \midrule
\multirow{9}{*}{\begin{tabular}[c]{@{}c@{}}Fine-Tuned LLMs\end{tabular}}& ChipNeMo~\cite{liu2023chipnemo} & 7B & \text{\texttimes} & 43.4 & - & \multicolumn{1}{c}{-} & 22.4 & - & - & \multicolumn{1}{c}{-} & - \\
 & RTLCoder-Mistral~\cite{liu2023rtlcoder} & 7B & \checkmark & 62.5 & 72.2 & \multicolumn{1}{c}{76.6} & 36.7 & 45.5 & 49.2 & \underline{96.6} & 48.3 \\
& RTLCoder-DS~\cite{liu2023rtlcoder} & 6.7B & \checkmark & 61.2 & 76.5 & {81.8} & 41.6 & 50.1 & 53.4 & {93.1} & 48.3 \\
& BetterV-CL~\cite{betterv} & 7B & \text{\texttimes} & 64.2 & 75.4 & \multicolumn{1}{c}{79.1} & 40.9 & 50.0 & 53.3 & \multicolumn{1}{c}{-} & - \\
& BetterV-DS~\cite{betterv} & 6.7B & \text{\texttimes} & 67.8 & 79.1 & \multicolumn{1}{c}{84.0} & 45.9 & 53.3 & 57.6 & \multicolumn{1}{c}{-} & - \\
& BetterV-CQ~\cite{betterv} & 7B & \text{\texttimes} & 68.1 & 79.4 & {84.5} & 46.1 & 53.7 & 58.2 & \multicolumn{1}{c}{-} & - \\
& CraftRTL-CL~\cite{liu2024craftrtl} & 7B & \text{\texttimes} & 78.1 & 85.5 & 87.8 & 63.1 & 67.8 & 69.7 & 93.9 & 52.9 \\
& CraftRTL-DS~\cite{liu2024craftrtl} & 6.7B & \text{\texttimes} & 77.8 & 85.5 & 88.1 & 65.4 & 70.0 & 72.1 & 84.3 & 58.8 \\
 \midrule
\multirow{4}{*}{CodeV-Verilog}& CodeV-Verilog-CL & 7B & \checkmark & {78.1} & 86.0 & \multicolumn{1}{c}{{88.5}} & 45.2 & 59.5 & 63.8 & \multicolumn{1}{c}{93.1} & 62.1 \\
& CodeV-Verilog-DS & 6.7B & \checkmark & {77.9} & \underline{88.6} & \multicolumn{1}{c}{{90.7}} & {52.7} & {62.5} & {67.3} & \multicolumn{1}{c}{89.7} & 55.2 \\
 & CodeV-Verilog-CQ & 7B & \checkmark & 77.6 & {88.2} & \multicolumn{1}{c}{{90.7}} & {53.2} & {65.1} & {68.5} & \multicolumn{1}{c}{93.1} & 55.2 \\
  & CodeV-Verilog-QC & 7B & \checkmark & \underline{80.1} & 87.9  & 90.5  & {59.2}  & {65.8}  & {69.1}  & \underline{96.6} & 51.7 \\
  \midrule
\multirow{4}{*}{CodeV-All}& CodeV-All-CL & 7B & \checkmark & 78.5 & 85.6  & 87.6  & 46.6  & 58.8  & 62.5  & \underline{96.6} & 55.2 \\
& CodeV-All-DS & 6.7B & \checkmark & 79.8 & 86.0  & 86.7  & 53.0  & 63.3  & 67.2  & \underline{96.6} & 51.7 \\
& CodeV-All-CQ  & 7B & \checkmark & 79.9 & 88.3  & \underline{91.1}  & 54.1  & 65.1  & 68.6  & 93.1 & 58.6 \\
& CodeV-All-QC  & 7B & \checkmark & \textbf{81.9} & \textbf{89.9}  & \textbf{92.0}  & {56.6}  & 67.9  & 71.4  & \underline{96.6} & 55.2 \\
\midrule
\multirow{2}{*}{CodeV-R1} & CodeV-R1-Distill & 7B & \checkmark & 76.2 & 85.6 & 87.0 & \underline{65.7} & \underline{76.8} & \underline{79.7} & - & \underline{75.8} \\
 & CodeV-R1 & 7B & \checkmark & 76.5 & 84.1 & 85.7 & \textbf{69.9} & \textbf{79.3} & \textbf{81.7} & - & \textbf{86.1} \\
\bottomrule
\end{tabular}%
}
\end{table*}

\begin{table*}[h]
\caption{Comparison of CodeV-R1 on RTLLM v2 and VerilogEval v2.}
\label{tab:ve2&rtllm2}
\centering
\footnotesize
\resizebox{0.95\textwidth}{!}{
\begin{tabular}{cccc ccc ccc ccc} 
\toprule
\multirow{2}{*}{Type} & \multirow{2}{*}{Model} & \multirow{2}{*}{\begin{tabular}[c]{@{}c@{}}Model\\ size\end{tabular}} & \multirow{2}{*}{\begin{tabular}[c]{@{}c@{}}Open\\ source\end{tabular}} & \multicolumn{3}{c}{VerilogEvalv2-SR (\%)} & \multicolumn{3}{c}{VerilogEvalv2-CC (\%)} & \multicolumn{3}{c}{RTLLM v2 (\%)} \\ 
& & & & pass@1 & pass@5 & pass@10 & pass@1 & pass@5 & pass@10 & pass@1 & pass@5 & pass@10 \\ 
\midrule
\multirow{8}{*}{\begin{tabular}[c]{@{}c@{}}Base LLMs\end{tabular}}
& GPT-4o & - & \text{\texttimes} & 64.1 & 73.7 & 76.2 & 57.6 & 66.1 & 69.0 & 56.5 & 70.3 & 75.2 \\ 
& DeepSeek-R1~\cite{r1} & 671B & \checkmark & \textbf{77.5} & \textbf{84.7} & \textbf{87.4} & \textbf{79.1} & \textbf{85.1} & \textbf{87.1} & \underline{64.7} & \underline{75.8} & \underline{79.7} \\ 
& DeepSeek-V3~\cite{dsv3} & 671B & \checkmark & 62.4 & 71.7 & 75.0 & 68.7 & 76.3 & 78.2 & 59.1 & 71.5 & 73.3 \\ 
& QWQ-32B~\cite{qwq32b} & 32B & \checkmark & 64.2 & 77.3 & 80.1 & 64.0 & 77.8 & \underline{80.9} & 52.9 & 68.0 & 71.2 \\ 
& DeepSeek-R1-Distill-Qwen-32B~\cite{r1} & 32B & \checkmark & 43.9 & 63.3 & 69.2 & 53.8 & 69.8 & 73.8 & 42.4 & 62.1 & 67.0 \\ 
& DeepSeek-R1-Distill-Qwen-7B~\cite{r1} & 7B & \checkmark & 0.6 & 2.2 & 3.5 & 2.0 & 7.0 & 11.3 & 0.0 & 0.0 & 0.0 \\ 
& Qwen2.5-Coder-32B-Instruct~\cite{qwen25coder} & 32B & \checkmark & 47.5 & 60.7 & 64.7 & 46.6 & 59.0 & 62.8 & 47.8 & 63.9 & 67.8 \\ 
& Qwen2.5-Coder-7B-Instruct~\cite{qwen25coder} & 7B & \checkmark & 31.3 & 49.3 & 54.6 & 30.5 & 46.8 & 52.0 & 36.1 & 52.4 & 57.6 \\ 
\midrule
& CodeV-R1-Distill & 7B & \checkmark & 65.2 & 75.2 & 77.5 & 65.5 & 75.6 & 78.2 & 57.2 & 71.9 & 77.1 \\ 
\multirow{-2}{*}{CodeV-R1} & CodeV-R1 & 7B & \checkmark & \underline{68.8} & \underline{78.2} & \underline{81.1} & \underline{69.9} & \underline{78.2} & \underline{80.9} & \textbf{68.0} & \textbf{78.2} & \textbf{81.7} \\ 
\bottomrule
\end{tabular}
}
\begin{tablenotes}
\item $^{\rm *}$ WSR: Specification-to-RTL. CC: Code Completion.
\end{tablenotes}
\end{table*}

HDLs play a pivotal role in processor chip design by enabling the creation of Register Transfer Level (RTL) code that connects natural language specifications with manufacturable chip layouts. As a critical determinant of functionality, performance, power efficiency, and production costs, HDL implementation currently consumes over 70\% of chip development cycles according to NVIDIA, highlighting the urgent need for automation solutions.

LLMs have shown revolutionary potential in software engineering, such as GitHub Copilot's 55\% efficiency improvement~\cite{peng2023impact}, yet their application to HDL generation remains suboptimal due to two challenges: 1) Data scarcity: Public code dataset contain 42$\times$ fewer Verilog samples (1.91M) than Python code (80.6M)~\cite{starcoder2}. Besides, there is almost no HDL-focused competition datasets like LeetCode or Codeforces in software; 2) Semantic disparity: HDLs demand precise low-level control, such as signal bit-width management, that creates significant abstraction gaps between specifications and implementations.

To tackle the above challenges, we present a multi-level summarization-based data synthesis approach and fine-tune general-purpose LLMs using the synthesized data to develop a series of HDL code generation models, referred to as CodeV~\cite{zhao2024codev}. CodeV implements the module generation component of the hardware design agent, which utilizes the LPCM for hardware design. 
Specifically, building on the insight that ``summarizing code to natural language is easier and more straightforward than generating code from natural language'', we develop a progressive abstraction technique that converts existing HDL code into high-quality natural language-code pairs, which effectively bridges the HDL-semantic gap. As shown in Table~\ref{tab:ve1&rtllm1}, this process yields 180k optimized training samples, enabling CodeV-Verilog to achieve 80.1\% pass@1 on VerilogEval-Machine~\cite{verilogeval}, surpassing previous SOTA open-source models RTLCoder~\cite{liu2023rtlcoder}. 

Based on CodeV, we make two key extensions:

1) To better align with real-world development workflows, we extended CodeV-Verilog into CodeV-All through the Chat-FIM-Tag supervised fine-tuning method. CodeV-All not only supports a wider range of languages, including Verilog and Chisel, and a broader set of tasks such as Chat and fill-in-the-middle (FIM), but it also delivers performance on VerilogEval that matches or even surpasses CodeV-Verilog (shown in Table~\ref{tab:ve1&rtllm1}), which was fine-tuned solely on Verilog. This makes the CodeV series the first set of open-source LLMs designed for multi-scenario HDL generation.

2) Inspired by the reasoning capabilities demonstrated in mathematical and software coding tasks, we proposed several innovations: a rule-based testbench generator that verifies predicted code against a golden reference, a round-trip data synthesis method that generates high-quality natural language-code pairs using only source code snippets as input, and adaptive DAPO, a fast version of DAPO~\cite{dapo} that dynamically adjusts the number of samples per step based on past sample discard rates. These components were integrated into a ``distill-then-RL'' two-stage training pipeline to develop CodeV-R1~\cite{codev-r1}, a reasoning-enhanced Verilog generation LLM that is capable of \textit{thinking} and test-time scaling. As shown in Table~\ref{tab:ve2&rtllm2}, CodeV-R1 achieves 68.6\% and 72.9\% pass@1 on VerilogEval v2 and RTLLM v2, respectively, outperforming previous state-of-the-art models by 12\% to 21\%, and matching or even exceeding the performance of the 671B DeepSeek-R1.


\subsection{Automated OS Configuration Optimization}

\begin{table}[t]
\centering
\caption{Results of automated OS configuration oaptimization by AutoOS compared with existing methods.}
\begin{tabular}{@{}cccc@{}}
\toprule
\multirow{2}{*}{Method} & \multicolumn{3}{c}{OS Configuration Task (UnixBench)}                                                                                                                                                        \\ \cmidrule(l){2-4} 
                        & \begin{tabular}[c]{@{}c@{}}PolyOS\\ on Sifive Unmatched\end{tabular} & \begin{tabular}[c]{@{}c@{}}Fedora\\ on Sifive Unmatched\end{tabular} & \begin{tabular}[c]{@{}c@{}}Ubuntu\\ on PC Machine\end{tabular} \\ \midrule
Default                 & 309                                                                  & 207                                                                  & 3885                                                           \\ \midrule
GPT-3.5                 & 283 (-8.5\%)                                                         & 194 (-6.3\%)                                                         & 3898 (+0.3\%)                                                  \\ \midrule
AutoOS                  & \textbf{335 (+8.4\%)}                                                & \textbf{260 (+25.6\%)}                                               & \textbf{4238 (+9.0\%)}                                         \\ \bottomrule
\end{tabular}
\label{tab:AutoOS}
\end{table}

The operating systems (OS) act as a crucial bridge between processors and higher-level software, playing a vital role in maximizing the performance of the processor chips. The widely used open-source OS Linux, designed to meet the diverse requirements of different application scenarios and processors, consists of over 20 million lines of code contributed by developers around the world, making it one of the most complex software projects to date. This vast codebase presents significant opportunities for optimization, and there is a pressing need to tailor or optimize the OS for specific processors and application scenarios to fully unleash the potential of the entire computer system.

However, customizing or optimizing an OS involves three main challenges. First, the complexity of the task is extremely high. Even just optimizing the OS kernel involves over 15,000 interdependent configuration options~\cite{oh2021finding,franz2021configfix}, which are beyond the capability of conventional optimization methods. Second, the cost of evaluating each configuration is high, as compiling, installing, and testing the OS can take up to 1 to 2 hours~\cite{xia2023comsa}, which limits the feasibility of data-driven methods like neural networks. Third, the optimization process is highly sensitive, where even a small error could prevent the OS from booting properly and make debugging extremely difficult.

To address these challenges, we leverage the LLM-guided performance feedback loop from our Software Design Agent to develop an automated OS configuration optimization method, AutoOS~\cite{chen2024autoos}, which can generate optimized kernel configurations without manual intervention and surpass the performance achieved by hardware vendors' manual optimizations. 
To achieve this, we introduce an ``observe-prune-propose-act-correct'' feedback loop, which leverages the prior knowledge embedded in LLMs to eliminate irrelevant configuration options that do not contribute to performance optimization and might cause booting issues, significantly reducing the search space for customization. In just a few search iterations, approximately one day, the method can automatically generate custom-optimized operating system kernel configurations. Compared to manual expert optimization, this approach can boost performance by as much as 25.6\%, as shown in Table~\ref{tab:AutoOS}.

\begin{table*}[t]
\caption{Comparison of our automated tensor program transcompiler against state-of-the-art LLMs on different transcompilation directions. 
Refer to the original paper~\cite{dong2025qimengxpilertranscompilingtensorprograms} for more details and complete results .}
\label{tab:qimeng-xpiler}
\centering
\footnotesize
{%
\begin{tabular}{@{}c|c|cccc|cccc@{}}
\toprule
\multirow{2}{*}{Source-Target} & \multirow{2}{*}{Method}             & \multicolumn{4}{c|}{Compilation Accuracy (\%)}                    & \multicolumn{4}{c}{Computation Accuracy (\%)}                   \\ \cmidrule(l){3-10} 
                               &                                     & CUDA C         & BANG C         & HIP            & C With VNNI    & CUDA C         & BANG C        & HIP            & C With VNNI   \\ \midrule
\multirow{3}{*}{CUDA C}        & GPT-4 & -              & 50.6           & 97.0           & 84.5           & -              & 7.7           & 96.4           & 30.4          \\
                               & OpenAI o1                           & -              & 51.8           & 98.2           & 85.1           & -              & 48.2          & 98.2           & 55.4          \\
                               & QiMeng-Xpiler                       & \textbf{-}     & \textbf{100.0} & \textbf{100.0} & \textbf{100.0} & -              & \textbf{91.7} & \textbf{100.0} & \textbf{95.2} \\ \midrule
\multirow{3}{*}{BANG C}        & GPT-4                               & 69.0           & -              & 66.1           & 23.8           & 6.5            & -             & 6.5            & 13.1          \\
                               & OpenAI o1                           & 71.4           & -              & 97.0           & 41.7           & 10.1           & -             & 7.7            & 23.2          \\
                               & QiMeng-Xpiler                       & \textbf{100.0} & -              & \textbf{100.0} & \textbf{100.0} & \textbf{95.8}  & -             & \textbf{97.0}  & \textbf{95.2} \\ \midrule
\multirow{3}{*}{HIP}           & GPT-4                               & 97.0           & 35.1           & -              & 85.1           & 97.0           & 5.4           & -              & 24.4          \\
                               & OpenAI o1                           & 98.8           & 42.3           & -              & 88.7           & 98.2           & 9.0           & -              & 30.4          \\
                               & QiMeng-Xpiler                       & \textbf{100.0} & \textbf{100.0} & -              & \textbf{100.0} & \textbf{100.0} & \textbf{86.9} & -              & \textbf{96.4} \\ \midrule
\multirow{3}{*}{C With VNNI}   & GPT-4                               & 81.5           & 41.7           & 74.7           & -              & 14.3           & 6.0           & 12.5           & -             \\
                               & OpenAI o1                           & 87.5           & 55.4           & 97.0           & -              & 51.2           & 10.7          & 96.4           & -             \\
                               & QiMeng-Xpiler                       & \textbf{100.0} & \textbf{99.4}  & \textbf{100.0} & -              & \textbf{98.2}  & \textbf{88.7} & \textbf{99.4}  & -             \\ \bottomrule
\end{tabular}
}
\end{table*}

\subsection{Automated Compiler Tool-Chain Design}

Compilers for modern processors are responsible for two fundamental tasks: 1) accurately and efficiently translating precise and unambiguous programming languages corresponding to the processor’s instruction set, i.e., translation; and 2) constructing compilation optimization sequences within a vast, high-dimensional optimization space, i.e., optimization. 
Currently, AI techniques in compilers are mainly focused on improving the optimization sequences within the high-dimensional space of existing compiler frameworks, also known as the Phase Ordering Problem. Yet they struggle to generate an end-to-end compiler that handles both two fundamental tasks for processors.

To address the long-term goal of creating an end-to-end compiler capable of both translation and optimization tasks, we have explored the automated compiler tool-chain design methods based on the Software Design Agent, investigating two different approaches: 1) automatically generating compiler backend code. Building upon existing architectures such as LLVM, we construct compiler backend datasets  ComBack~\cite{zhong2024comback} and fine-tune the LLMs to improve and fully exploit LLMs' comprehension ability for compiler backend code  VEGA~\cite{zhong2025vega}. As a result, we successfully generated compiler backend code tailored to a specific processor with an accuracy rate exceeding 70\%, with explicit confidence scores highlighting critical regions requiring minimal manual refinement. This approach promises to revolutionize conventional backend development workflows.
2) Using LLM as an end-to-end compiler. We discover that the translation task of compilers shares significant similarities with natural language translation, an area where LLMs excel. This suggests that LLMs have the potential to revolutionize compiler construction to act as a real compiler. However, since natural languages are inherently ambiguous while programming languages have precise semantics defined by grammar, directly applying LLMs to translation tasks leads to suboptimal results. For instance, using GPT-4 for translating C language to RISC-V assembly yields an accuracy rate below 50\%, with complex functions performing near zero. 
Therefore, we proposed an end-to-end neural compiler method~\cite{zhang2024introducing} based on the Software Design Agent. 
This method combines grammar information from programming languages and compiler domain knowledge to guide the generation of specialized LLMs. On one hand, data augmentation techniques guided by compiler expertise were used to create high-quality datasets to fine-tune LLMs. On the other hand, we leverage the program's grammar information during the inference stage for LLMs tailored to the specific translation task. This combination enabled us to achieve over 99\% accuracy for C language translation on the ExeBench~\cite{armengol2022exebench} dataset and successfully compile code from real-world datasets like AnsiBench~\cite{ansibench} and CoreMark~\cite{coremark}, confirming the feasibility of this approach. Going forward, we will continue to refine how to enhance the performance of the Software Design Agent based on LPCM used directly as end-to-end compilers.

\subsection{Automated Tensor Program Transcompiler}

Contemporary LLMs, including prominent examples like GPT and DeepSeek, exhibit deep dependencies on NVIDIA's CUDA ecosystem. This reliance encompasses both vendor-provided libraries such as cuBLAS~\cite{cublas}, cuDNN~\cite{cudnn}, TensorRT~\cite{NVTRT}, and community-developed kernels such as FlashAttention-v1~\cite{dao2022flashattention}, FlashAttention-v2~\cite{dao2023flashattention}, FlashAttention-v3~\cite{shah2024flashattention}. 
Even the domestic open-source LLM DeepSeek~\cite{liu2024deepseek} has also developed tailored acceleration libraries like FlashMLA~\cite{flashmla2025} and DeepGEMM~\cite{deepgemm2025} for NVIDIA GPUs. 
However, the software ecosystem for domestic AI chips faces significant fragmentation, as different chip manufacturers develop their own independent operator libraries. This makes it challenging to unify the software ecosystems across domestic AI chips, hindering the widespread adoption of these chips.

To address this challenge, we have developed an automated tensor program tanscompiler, QiMeng-Xpiler~\cite{dong2025qimengxpilertranscompilingtensorprograms}, based on the Software Design Agent, enabling ``Write Once, Run Anywhere'' across different AI chips, including both NVIDIA GPUs and domestic AI Chips. 
The key is that the program translation process is automatically conducted as a series of neural-symbolic transformation passes based on the function adaptation feedback loop, where LLMs generate high-level program sketches, and the incorrect code details are repaired by small-scale symbolic synthesis.
Meanwhile, the optimal transformation passes are identified via hierarchical auto-tuning based on the performance optimization feedback loop.
Specifically, we combine inter-pass Monte Carlo Tree Search~\cite{browne2012survey} for optimal transformation sequencing and intra-pass constraint-based auto-tuning of critical tuning parameters, such as memory tiling configurations.  
Ultimately, our solution enables an automated tensor program transcompiler across various processors like Nvidia GPUs~\cite{NVGPU}, Cambricon MLU~\cite{MLU}, AMD MI accelerators~\cite{AMDGPU}, Intel DLBoost~\cite{DLBoost}, and programming models like SIMT, SIMD. In real-world applications such as LLMs, experiments on those $4$ diverse processors demonstrate that QiMeng-Xpiler correctly translates different tensor programs at the accuracy of 95\% on average, as shown in Table~\ref{tab:qimeng-xpiler}. 


\begin{table}[t]
\centering
\label{tab:qimeng-gemm}
\caption{Performance comparison of matrix multiplication methods across different hardware platforms C910(GFLOPS), NVIDIA RTX4070(TFLOPS), NVIDIA A100(TFLOPS). Speedup ratios for QiMeng-GEMM are calculated against OpenBLAS (C910) and cuBLAS (RTX 4070, A100). The A100 and RTX4070 GPU utilizes CUDA cores. }
\resizebox{0.95\linewidth}{!}{
\begin{tabular}{@{}c|c|rrr@{}}
    \toprule
    \multirow{2}{*}{Hardware} & \multirow{2}{*}{Method} & \multicolumn{3}{c}{Dimension (M = K = N)} \\ \cmidrule{3-5}
                              &                         & \multicolumn{1}{c}{1024} & \multicolumn{1}{c}{2048} & \multicolumn{1}{c}{4096} \\
    \midrule
                              & GPT-4o~\cite{gpt4o}                  & 0.14                     & 0.10                     & 0.09                     \\
C910                          & Claude 3.5 Sonnet~\cite{claude35}       & 2.64                     & 1.56                     & 0.74                     \\
(RISC-V)                      & OpenBLAS~\cite{openblas}                & 5.01                     & 5.11                     & 4.85                     \\
                              & QiMeng-GEMM             & 9.91(1.98×)              & 10.08(1.97×)             & 10.23(2.11×)             \\
    \midrule
                              & GPT-4o                  & 1.77                     & 1.78                     & 1.65                     \\
RTX 4070                      & Claude 3.5 Sonnet       & 1.71                     & 1.79                     & 1.61                     \\
(NVIDIA)                      & cuBLAS~\cite{cublas}                  & 10.79                    & 12.77                    & 12.78                    \\
                              & QiMeng-GEMM             & 11.47(1.06×)             & 13.31(1.04×)             & 14.16(1.11×)             \\
    \midrule
                              & GPT-4o                  & 4.19                     & 4.27                     & 4.71                     \\
A100                          & Claude 3.5 Sonnet       & 4.64                     & 5.33                     & 5.27                     \\
(NVIDIA)                      & cuBLAS                  & 16.26                    & 17.20                    & 18.97                    \\
                              & QiMeng-GEMM             & 12.61(0.77×)             & 16.17(0.94×)             & 18.27(0.96×)             \\
    \bottomrule
\end{tabular}
}

\end{table}

\subsection{Automated High-Performance Library Generation}

Leading hardware vendors, such as NVIDIA, ARM, Intel, AMD, and Cambricon, invest heavily in manually optimized libraries to extract peak performance from their processors. These expert-crafted solutions demand intimate knowledge of microarchitecture details, requiring careful parallelization of computations and memory operations, often implemented in vendor-specific languages or assembly code. 
While delivering exceptional performance, this manual optimization paradigm fundamentally lacks scalability and portability across different hardware architectures.

\begin{table}[t]
\centering
\caption{Performance comparison of GEMM and convolution operations across different hardware platforms, measured in GFLOPS (K1, A76) and TFLOPS (A100). The A100 GPU utilizes Tensor Cores, with speedup ratios (in parentheses) for QiMeng-TensorOp calculated against OpenBLAS (K1, A76) and cuBLAS/cuDNN (A100).}
\label{tab:qimeng-tensorop}
\resizebox{0.95\linewidth}{!}{
\begin{tabular}{@{}c|c|rrr@{}}
\toprule
\multirow{2}{*}{Hardware} & \multirow{2}{*}{Method} & \multicolumn{3}{c}{Matrix Multiplication (M=K=N)} \\
\cmidrule{3-5}
                          &                         & \multicolumn{1}{c}{1024} & \multicolumn{1}{c}{2048} & \multicolumn{1}{c}{4096} \\
\midrule
\multirow{3}{*}{\begin{tabular}[c]{@{}c@{}}K1\\(RISC-V)\end{tabular}} 
                          & DeepSeek-V3~\cite{liu2024deepseek}             & 0.33                      & 0.31                      & 0.23                      \\
                          & OpenBLAS~\cite{openblas}                & 4.19                      & 4.46                      & 4.76                      \\
                          & QiMeng-TensorOp         & 9.74(2.32×)               & 10.29(2.31×)              & 11.74(2.47×)              \\
\midrule
\multirow{3}{*}{\begin{tabular}[c]{@{}c@{}}A76\\(ARM)\end{tabular}}     
                          & DeepSeek-V3             & 0.04                      & 0.04                      & 0.04                      \\
                          & OpenBLAS                & 31.25                     & 33.48                     & 34.27                     \\
                          & QiMeng-TensorOp         & 35.70(1.14×)              & 36.77(1.10×)              & 37.31(1.09×)              \\
\midrule
\multirow{3}{*}{\begin{tabular}[c]{@{}c@{}}A100\\(NVIDIA)\end{tabular}}  
                          & DeepSeek-V3             & 17.74                     & 17.31                     & 18.76                     \\
                          & cuBLAS~\cite{cublas}                  & 246.10                    & 292.20                    & 298.44                    \\
                          & QiMeng-TensorOp         & 262.05(1.06×)             & 290.86(1.00×)             & 293.44(0.98×)             \\
\midrule\midrule
\multirow{4}{*}{Hardware} & \multirow{4}{*}{Method} & \multicolumn{3}{c}{Shape of Feature Map (N, C, H, W)} \\
                          &                         & \multicolumn{3}{c}{Shape of Filter (K, C, R, S)} \\ \cmidrule{3-5}

                          &                         & \multicolumn{1}{c}{(64,64,56,56)} & \multicolumn{1}{c}{(64,128,56,56)} & \multicolumn{1}{c}{(32,512,14,14)} \\
                          &                         & \multicolumn{1}{c}{(64,64,3,3)}   & \multicolumn{1}{c}{(128,128,3,3)}  & \multicolumn{1}{c}{(512,512,3,3)}  \\
\midrule
\multirow{3}{*}{\begin{tabular}[c]{@{}c@{}}K1\\(RISC-V)\end{tabular}}  
                          & DeepSeek-V3             & 0.01                      & 0.01                      & 0.01                      \\
                          & OpenBLAS                & 6.33                      & 6.51                      & 7.31                      \\
                          & QiMeng-TensorOp         & 6.55(1.03×)               & 8.08(1.24×)               & 8.96(1.23×)               \\
\midrule
\multirow{3}{*}{\begin{tabular}[c]{@{}c@{}}A76\\(ARM)\end{tabular}}     
                          & DeepSeek-V3             & 0.06                      & 0.03                      & 0.05                      \\
                          & OpenBLAS                & 12.97                     & 19.33                     & 27.92                     \\
                          & QiMeng-TensorOp         & 28.82(2.22×)              & 30.84(1.60×)              & 32.98(1.18×)              \\
\midrule
\multirow{3}{*}{\begin{tabular}[c]{@{}c@{}}A100\\(NVIDIA)\end{tabular}}  
                          & DeepSeek-V3             & 14.77                     & 20.51                     & 14.79                     \\
                          & cuDNN~\cite{cudnn}                   & 117.96                    & 120.59                    & 136.63                    \\
                          & QiMeng-TensorOp         & 116.73(0.99×)             & 121.48(1.01×)             & 125.71(0.92×)             \\
\bottomrule
\end{tabular}
}
\end{table}

To address these challenges, in addition to leveraging existing software ecosystems through the aforementioned automated tensor program transcompiler, we pioneer an automated approach called QiMeng-GEMM~\cite{zhou2025QiMeng} based on Software Design Agent for generating high-performance libraries with matrix multiplication, i.e. GEMM, as our primary target due to its central role in LLMs~\cite{llama2,liu2024deepseek}, deep learning~\cite{devlin2019bert, dosovitskiy2020image}, and scientific computing~\cite{feng2021egemm}. 
The proposed QiMeng-GEMM is the first to automatically generate high-performance GEMM code by exploiting LLMs. 
Specifically, we have abstracted common GEMM optimization methods and hardware architecture features, and created a set of general meta-prompts for LLMs to generate high-performance matrix multiplication operators. These meta-prompts enable LLMs to understand and implement optimization goals by capturing the architectural features of different platforms. We then integrate the performance feedback loop in the Software Design Agent with Tree of Thoughts~\cite{yao2023tree} (ToT) techniques to systematically explore optimization primitive combinations. 
This allows us to explore all possible optimization sequences generated by the meta-prompts, thus enabling the generation of high-performance matrix multiplication operators that are tailored to different hardware architecture features.

Further extending our LLM-based automation framework, we propose QiMeng-TensorOp~\cite{zhang2025qimengtensoropautomaticallygeneratinghighperformance}, the first approach to automatically generate high-performance tensor operators with hardware primitives by leveraging LLMs. We develop structured hardware-intrinsic optimization prompts and a knowledge-guided workflow, enabling LLMs to comprehend platform-specific architectures and optimization strategies. To optimize the generated operators, we design an LLM-guided Monte Carlo Tree Search (MCTS) algorithm, which effectively enhances the efficiency and performance of tuning primitive-level tensor operators on specific hardware.

We further propose QiMeng-Attention, the first hardware-aware automated framework for cross-platform Attention operator generation. We propose an LLM-friendly Thinking Language (LLM-TL) to help LLMs decouple the generation of high-level optimization logic and low-level implementation on GPU, and enhance LLMs' understanding of the attention operator. Along with a 2-stage reasoning workflow, TL-Code generation and translation, the LLMs can automatically generate FlashAttention implementation on diverse GPUs, establishing a self-optimizing paradigm for generating high-performance attention operators in attention-centric algorithms.

We have validated these approaches on diverse platforms such as the Xuantie C910 development board~\cite{chen2020xuantie}, MuseBook (K1)~\cite{musebook}, ARM A76~\cite{arm_a76}, and NVIDIA GPUs (RTX 4070~\cite{nvidia_rtx4070}, RTX 8000~\cite{nvidia_rtx8000}, T4~\cite{nvidia_t4} and A100~\cite{nvidia_a100}), see Table \ref{tab:qimeng-gemm}, Table \ref{tab:qimeng-tensorop} and Table \ref{tab:qimeng-attention}. On the RISC-V platform, the high-performance matrix multiplication operator generated by QiMeng-GEMM and QiMeng-TensorOp achieves up to 211\% and 251\% of OpenBLAS's performance, respectively. 
On the NVIDIA platform, they reach up to 115\% and 124\% of cuBLAS's performance, respectively. 
Compared to conventional LLM prompt methods, our approach significantly improves the performance of the generated code and boosts development efficiency.
To validate the performance of the Qimeng-Attention, we conducted experiments across various NVIDIA hardware architectures.
On the NVIDIA T4 platform and NVIDIA RTX8000 platform, the high-performence attention operator generated by Qimeng-Attention consistently achieves superior performance metrics compared to all four implementations.

\begin{table}[t]
\centering
\caption{Performance (TFLOPS) comparison across attention operators, NVIDIA GPUs (T4, RTX 8000, A100) under the configuration of head dimension 128, sequence length 2048, batch size 8, head number 16, GQA groups 8 and without causal mask. Speedup ratios are calculated against the PyTorch implementation of DeepSeek-V3.}
\label{tab:qimeng-attention}
\resizebox{0.95\linewidth}{!}{
\begin{tabular}{@{}c|c|rrr@{}}
    \toprule
    \multirow{2}{*}{Hardware} & \multirow{2}{*}{Method} & \multicolumn{3}{c}{Variant of Attention} \\ 
    \cmidrule{3-5}
                              &                         & \multicolumn{1}{c}{MQA} & \multicolumn{1}{c}{GQA} & \multicolumn{1}{c}{MQA} \\
    \midrule
\multirow{5}{*}{\begin{tabular}{c}T4\\(NVIDIA)\end{tabular}} 
                              & cuDNN~\cite{cudnn}                   & 12.95                   & 13.02                   & 13.03                   \\
                              & FlexAttention~\cite{dong2024flex}           & 14.83                   & 14.95                   & 14.64                   \\
                              & Flash Attention v1~\cite{dao2022flashattention}      & 10.95                   & 10.95                   & 11.01                   \\
                              & DeepSeek-V3~\cite{liu2024deepseek}             & 6.11                    & 3.97                    & 5.99                    \\
                              & QiMeng-Attention        & 18.59(3.04×)            & 18.82(4.74×)            & 18.14(3.03×)            \\
    \midrule
\multirow{5}{*}{\begin{tabular}{c}RTX 8000\\(NVIDIA)\end{tabular}} 
                              & cuDNN                   & 32.2                   & 32.1                   & 31.2                   \\
                              & FlexAttention           & 33.2                   & 33.4                   & 33.5                   \\
                              & Flash Attention v1      & 21.2                   & 21.1                   & 21.3                   \\
                              & DeepSeek-V3             & 13.4                   & 8.8                    & 13.2                   \\
                              & QiMeng-Attention        & 44.9(3.35×)            & 43.3(4.92×)            & 43.4(3.29×)            \\
    \midrule
\multirow{5}{*}{\begin{tabular}{c}A100\\(NVIDIA)\end{tabular}} 
                              & cuDNN                   & 190.0                  & 189.6                  & 189.9                  \\
                              & FlexAttention           & 143.2                  & 143.5                  & 143.5                  \\
                              & Flash Attention v2~\cite{shah2024flashattention}      & 208.2                  & 200.0                  & 200.7                  \\
                              & DeepSeek-V3             & 52.4                   & 23.1                   & 38.4                   \\
                              & QiMeng-Attention        & 201.1(3.84×)           & 186.2(8.06×)           & 187.6(4.89×)           \\
    \bottomrule
\end{tabular}
}

\end{table}
\section{Related Work}
\label{sec6}
\subsection{Automated Chip Design}

Automating the design of processor chips is one of the key challenges in computer science. Early EDA tools, based on predefined rules and Boolean logic, enabled the automation of specific design steps such as logic synthesis, placement, and routing. Later, researchers introduced automated design methodologies based on domain-specific languages (DSLs), HLS, and DSE, etc. With advancements in AI, the automated design of processor chips has evolved into a more intelligent, data-driven phase by leveraging AI technologies. Techniques like random forests, RL, and GNNs are now being applied to enhance the EDA workflow in tasks such as automated performance evaluation, placement, and routing. However, these approaches primarily use AI to optimize the efficiency or performance of existing EDA processes without altering the fundamental processor chip design flow. In recent years, the concept of fully automated processor chip design has become a prominent research area, with approaches utilizing RL, random forests, and LLMs to design processor chips from functional requirements or design specifications without human effort. Nonetheless, current efforts still face challenges in improving the scale and accuracy of designed processor chips.

\subsubsection{EDA-based Automated Chip Design}
The conventional design flow based on EDA tools can be roughly categorized into three stages: logic design, circuit design, and physical design~\cite{chen2022chip}. With AI technology advancements, AI-based methods have been integrated into these three stages. The primary objective of these methods is to enhance specific steps within the conventional flow, thereby improving flow efficiency and design performance, instead of fundamentally altering the conventional flow~\cite{chen2022chip,he2023chip}.

The logic design stage aims to generate a hardware description, represented by HDLs such as Verilog and VHDL. This is achieved by either manually programming based on functional requirements or utilizing HLS tools based on hardware functionalities described in high-level programming languages such as C, C++, or SystemC. The former approach simplifies the design flow through hardware abstraction. For instance, Nurvitadhi et al.~\cite{Nurvitadhi2011} propose an automated transaction-to-pipeline transcompilation methodology. The ASSIST framework~\cite{Liu2019} supports RISC architecture design via micro-operation languages but lacks control over pipeline optimization. TL-Verilog~\cite{Hoover2017} partitions combinational logic through temporal abstraction but exhibits deficiencies in data hazard detection. Languages such as BSV~\cite{Nikhil2004} and Koika~\cite{Bourgeat2020} facilitate formal verification but enforce single-cycle rule execution without dynamic scheduling. The latter approach generates hardware descriptions from C/C++. For example, Rokicki et al.~\cite{rokicki2019you} generate processor cores from C++ while requiring manual handling of bypass logic. Josipović et al.~\cite{josipovic2018dynamically} introduce dynamic scheduling to optimize pipeline performance, while Dahlia~\cite{nigam2020predictable} leverages affine types to ensure predictability in statically scheduled accelerators. However, these conventional methods rely on formal language template conversion, which incurs high learning costs and constrains design spaces. Thus, recent advancements employ AI algorithms for rapid estimation of quality, performance, and timing to enhance HLS efficiency. For example, Zhao et al.~\cite{zhao2019machine} utilize linear regression and Artificial Neural Networks (ANNs) to predict routing congestion in HLS. Makrani et al.~\cite{makrani2019pyramid} propose a neural network (NN)-based approach to predict resource utilization and performance on specific field programmable gate arrays (FPGAs), thereby improving the efficiency of DSE. Ferianc et al.~\cite{ferianc2020improving} employ Gaussian processes for latency estimation to optimize accelerator configuration selection.

The circuit design stage, also known as logic synthesis, aims to transform hardware descriptions into gate-level circuits, i.e., netlists. During this stage, Boolean expressions and logical structures are optimized based on specified process libraries to achieve minimal logical expressions and netlists. LSOracle~\cite{neto2019lsoracle} employs Deep Neural Networks (DNNs) to intelligently differentiate circuit modules, dynamically selecting the most effective optimizers between And-Inverter Graph (AIG) and Majority-Inverter Graph (MIG) representations. Haaswijk et al.~\cite{haaswijk2018deep} and Zhu et al.~\cite{zhu2020exploring} reformulate conventional logic optimization as Markov Decision Processes (MDPs), developing deep RL systems that utilize Graph Convolutional Neural Networks (GCNN) as policy networks. Hosny et al.~\cite{hosny2020drills} implement an Advantage Actor-Critic (A2C) RL algorithm to minimize area while adhering to strict timing constraints. Deep-PowerX~\cite{pasandi2020deep} establishes an accurate error prediction model using DNNs to evaluate approximation circuit errors, enabling significant reductions in dynamic power while maintaining acceptable accuracy thresholds.

The physical design stage aims to generate layouts through placement, clock tree synthesis (CTS), and routing. During the placement stage, AI techniques are primarily employed to produce superior layouts. For example, Google formalizes placement as a sequential decision-making problem that can be addressed using RL, resulting in human-competitive placement within 6 hours~\cite{mirhoseini2020chip}.  During the CTS stage, AI techniques play a crucial role in optimizing clock tree structures and predicting performance. Lu et al.~\cite{lu2019gan} integrate generative adversarial networks (GANs) with RL to minimize clock skew and total clock tree length through topology prediction and optimization. Nagaria and Deb~\cite{nagaria2020designing}, along with Kwon et al.~\cite{kwon2018transient}, utilized convolutional neural networks (CNNs) and DNNs, respectively, to predict critical CTS parameters, including gating cell counts, buffer distributions, and wireload characteristics, thereby significantly enhancing the quality of synthesis. During the routing stage, AI techniques contribute to routing prediction and estimation. He and Bao~\cite{he2020circuit} apply RL to train agents for autonomous decision-making regarding spatial search strategies during routing optimization, dynamically selecting optimal neighboring nodes to enhance design quality. Liang et al.~\cite{liang2017empirical} and Alawieh et al.~\cite{alawieh2020high} model routing congestion prediction as image-to-image translation tasks using CNNs and conditional GANs, respectively, achieving high-precision hotspot predictions to guide routing optimization.

\subsubsection{Fully Automated Chip Design}
In recent years, fully automated chip design has emerged as a prominent research focus, particularly in automating front-end chip design directly from functional specifications and design specifications. For instance, PrefixRL~\cite{roy2021prefixrl} applies RL to synthesize circuits of approximately 100 gates; Chen et al.~\cite{chen2020circuit} uses random forests to design 200-gate circuits; and Rai et al.~\cite{rai2021logic} employ ensemble learning techniques to automatically design circuits with up to 2500 gates. However, these efforts are limited in scale and fall short on the precision requirements for complex circuits like CPUs. Moreover, the limited validation through small test cases in existing approaches fails to ensure the robustness needed for industrial chip fabrication.

Additionally, academia and industry have begun exploring LLM-based chip logic design, leveraging the natural language comprehension capabilities of LLMs to enable end-to-end chip generation. These approaches generally fall into two categories: foundation models and generation frameworks. Foundation models focus on generating HDL code for module-level designs. Among them, DAVE~\cite{pearce2020dave} fine-tunes GPT-2~\cite{gpt2} to produce Verilog code with 94.8\% accuracy. VeriGen~\cite{thakur2024verigen} improves generation quality by training on a hybrid dataset of source code and textbooks. ChipNeMo~\cite{liu2023chipnemo} addresses data scarcity through domain-adaptive pretraining of Llama2~\cite{llama2} on internal datasets from NVIDIA. RTLCoder~\cite{liu2023rtlcoder} creates a high-quality SFT dataset by evolving language prompts, generating code with GPT-3.5, and manually refining it, ultimately surpassing the original model's performance. The Large Circuit Model (LCM)~\cite{largecircuitmodels} enables feature extraction to accelerate SAT solving and aid in logic synthesis and equivalence checking. Notably, NVIDIA’s proprietary CraftRTL model~\cite{liu2024craftrtl} previously led the field, achieving 53.1\% accuracy on the RTLLM benchmark. 


Generation frameworks capitalize on LLMs’ planning and reflection capabilities to decompose complex hardware design tasks into manageable subtasks, iteratively refining the output based on environmental feedback.  For example, ChipGPT~\cite{chang2023chipgpt} implements a four-phase pipeline—prompt generation, initial Verilog generation, correction and optimization, and best-design selection—which has proven effective for CPU design. AutoChip~\cite{thakur2023autochip} reduces syntax errors by integrating compile-time diagnostics into the generation loop, boosting test pass rates by 24.2\%. Chip-Chat~\cite{chip-chat} uses dialogue-based task decomposition and human feedback to design an 8-bit accumulator microprocessor. RTLFixer~\cite{tsai2024rtlfixer} introduces retrieval-augmented generation to iteratively repair syntax errors in RTL code. ChatEDA~\cite{wu2024chateda} automates the RTL-to-GDSII pipeline through task planning, script generation, and execution via LLMs.

Despite these advances, existing LLM-based techniques remain constrained to small-scale module synthesis, document retrieval, and syntax correction. The generation of complex designs still heavily relies on human involvement, limiting the practical application of current methods in meeting stringent design cost and correctness requirements.

\subsection{Automated Software Design}

The field of automated software design encompasses two primary research directions: 1) functional adaptation, which enables functional-correctness automated cross-platform/cross-language software migration, and 2) performance optimization, which improves computational efficiency while maintaining platform/language compatibility. 

\subsubsection{Automated Software Adaptation}

In the area of foundational software function adaptation, typical applications include automated compiler tool-chain design and automated tensor program transcompiler. Existing research methods can be broadly categorized into three approaches: conventional rule-based automation, SMT-based symbolic synthesis, and data-driven methods.

Conventional rule-based methods rely on experts manually defining transformation rules for ASTs and achieving program translation through pattern matching. Notable works include the FCUDA~\cite{papakonstantinou2013efficient} framework, which automates the translation of CUDA to FPGA by defining rules for data communication, computation optimization, and parallel mapping; AMD's HIPIFY tool~\cite{HIPIFY}, which automates the migration of CUDA code from Nvidia GPUs to AMD GPUs; and source-to-source compilers like cxgo~\cite{cxgo} and C2Rust~\cite{C2Rust} that follow similar approaches. 
However, the architectural differences between platforms and languages make it extremely difficult to manually design efficient translation rules, and this approach struggles to handle the exponentially growing combinatorial space of foundational software adaptation.

SMT-based symbolic synthesis methods generate semantically equivalent target code through domain-specific languages or input-output examples. Key works include the Chlorophyll~\cite{phothilimthana2014chlorophyll} framework, which defines a domain-specific language for the GreenArrays GA144~\cite{GA144} architecture, breaking the translation process into subproblems such as partitioning, layout, and code generation, and using symbolic synthesis to produce functionally equivalent code. 
The FACC~\cite{woodruff2022bind} framework also uses program synthesis based on I/O examples, generating adaptation layer code to bridge the semantic gap between conventional Fourier transform programs and dedicated hardware accelerators. While these methods rely heavily on SMT solvers for constraint solving, they have two main limitations: First, search-based solvers (like Z3) are difficult to scale to general large-scale programs, and second, manual specification of input constraints adds significant engineering overhead.

Data-driven methods have rapidly advanced in recent years, utilizing vast amounts of data to train neural networks for software function migration and adaptation. Early examples include the neural compiler which uses the Transformer model for end-to-end compilation from C to x86 assembly~\cite{armengol2021learning}, and Meta AI’s Transcoder~\cite{roziere2020unsupervised} framework which uses back-translation learning to achieve cross-language translation between C, Python, and JAVA. There have also been some significant breakthroughs in recent years. CodeXGLUE~\cite{lu2021codexglue} is a code intelligence benchmarking system, and their CodeGPT~\cite{lu2021codexglue} model has significantly improved code generation capabilities; BabelTower~\cite{wen2022babeltower} significantly optimizing the C to CUDA translation via the large-scale datasets and proposed metrics; and QiMeng-GEMM~\cite{zhou2025QiMeng} which introduces the first automatically framework for generating high-performance matrix multiplication code without human effort. 
Although these methods have made remarkable progress, the correctness of the generated code is not yet fully assured, requiring manual verification and correction, which remains a key challenge in the field.

\subsubsection{Automated Software Optimization}

In the field of foundational software performance optimization, typical application scenarios include the automatic OS configuration optimization and the automatic high-performance library generation. Existing research methods can be broadly categorized into three approaches: conventional expert manual optimization, online learning-based search methods, and LLM-guided efficient search methods.

Conventional expert manual optimization methods rely on domain experts who use their experience to develop optimization strategies. Notable works include: the default operating system configuration options provided by processor manufacturers to maximize hardware performance; high-performance libraries such as cuBLAS~\cite{cublas}, cuDNN~\cite{cudnn}, and TensorRT~\cite{NVTRT}, manually optimized by thousands of engineers at Nvidia; and the FlashAttention series libraries~\cite{dao2022flashattention,dao2023flashattention,shah2024flashattention}, which were meticulously optimized by expert community developers. 
However, with the rapid evolution of algorithm models and the diversification of hardware architectures, this method which heavily depends on manual optimization, is increasingly facing significant engineering cost challenges.

Online learning-based search methods leverage AI algorithms such as machine learning and deep learning to automatically explore the optimization space. In OS configuration optimization, HiPerBOt~\cite{menon2020auto} uses Bayesian optimization to adjust application and platform configuration parameters to improve performance; Wayfinder~\cite{jung2021wayfinder} system applies Bayesian optimization to automatically optimize over 200 network configuration parameters in the operating system kernel. 
In high-performance library generation and optimization, AutoTVM~\cite{chen2018learning} uses XGBoost~\cite{chen2016xgboost} to train cost models, minimizing the overhead of hardware performance testing, and automates tuning within a search space defined by expert-specified scheduling templates. 
Ansor~\cite{zheng2020ansor} generates additional candidate program templates based on expert rules and uses genetic algorithms~\cite{vikhar2016evolutionary} to efficiently explore the search space. Although these methods have automated certain aspects, they still rely on manually defined search spaces and face challenges such as the long time required for online tuning, limiting their widespread application.

LLM-guided efficient search methods, with their built-in expert knowledge, robust natural language understanding ability, and code comprehension capabilities, introduce a new paradigm for automated software optimization. The AutoOS~\cite{chen2024autoos} framework implements a ``observe-prune-prose-act-correct'' performance feedback loop, leveraging the prior knowledge of LLMs to extend the OS configuration optimization problem to the full space of about 15,000 Linux kernel configuration options. The researchers from the University of Science and Technology of China introduced the TLM~\cite{zhai2024enabling} framework, transforming the automatic tuning of high-performance programs into a probability generation problem based on LLMs, facilitating more efficient automatic search than random sampling through enhanced semantic understanding. 
These methods combine LLMs' prior knowledge to implement efficient search strategies that resemble those of human experts, and have become a significant development direction for foundational software performance optimization.

\section{Conclusion and Future Work}\label{sec7}

The conventional paradigm of processor chip design is confronting three fundamental challenges: physical constrains in fabrication technology, requirements of design resources, and growing diversity of ecosystems. To achieve automated processor chip design based on LLMs and AI technologies, this work proposes QiMeng, an innovative paradigm for fully automated hardware and software design for processors and chips. QiMeng establishes a domain-specific LPCM and further develops Hardware Design Agent and Software Design Agent by leveraging the powerful knowledge representation and inferencing capabilities of LPCM. Then the two agents are applied to various application scenarios in processor hardware/software design. Currently, several components of QiMeng have been applied in multiple applications, providing viable solutions for hardware/software design.

With the deeper convergence of AI technologies and EDA tools, automated processor chip design will evolve toward greater efficiency, generality, and intelligence. In future work, we will follow the roadmap of QiMeng to accomplish the top-down phase, then proceed to the bottom-up and iteration phases. Simultaneously, we will continue exploring the integration of reinforcement learning, continual learning, and evolutionary algorithms to further enhance the capabilities of QiMeng. 
For the Hardware Design Agent, we will explore combining the framework of LPCM-based automated HDL generation with verify-repair-driven feedback and performance-driven feedback, establishing the entire Hardware Design Agent which achieves both correctness and performance optimization. Additionally, we will investigate end-to-end design from functional specifications to transistor-level implementation, breaking through conventional Boolean logic and CMOS paradigms, while scaling up the design to achieve industrial-grade automated processor chip design. 
For the Software Design Agent, the current implementation primarily exploits the textual comprehension from LLMs. Future enhancements will integrate the graph-structured representation. Moreover, we will extend the agent's applicability to autonomous software migration, functional adaptation, and performance optimization for more foundational software. 
By realizing and continuously improving the capabilities of QiMeng, we aim to address increasingly diverse scenario demands, driving the entire processor chip domain toward intelligence and automation.

\bibliographystyle{IEEEtran}
\bibliography{references}

\end{document}